\documentclass[aps,pre,reprint,twocolumn,footinbib,showpacs,amsmath,amssymb,sort&compress,superscriptaddress]{revtex4-1}
\usepackage{amsmath,amssymb}
\usepackage{graphicx}
\usepackage{xcolor}
\usepackage{verbatim}
\usepackage{titlesec}
\usepackage{array}
\usepackage{mathrsfs}

\usepackage{tikz}

\newcommand{\triadia}{\,%
\mbox{\raisebox{-0.3\height}{\begin{tikzpicture}
  [scale=.07,auto=left,every node/.style={circle,scale=0.3,fill=black}]
  \node (n1) at (1,1) {1};
  \node (n2) at (5,8)  {2};
  \node (n3) at (9,1)  {3};

  \foreach \from/\to in {n1/n2,n2/n3,n3/n1}
    \draw (\from) -- (\to);
\end{tikzpicture}}}\,%
}
\newcommand{\twodia}{%
\,\mbox{\raisebox{-0.0\height}{\begin{tikzpicture}
  [scale=.07,auto=left,every node/.style={circle,scale=0.3,fill=black}]
  \node (n1) at (1,1) {1};
  \node (n2) at (9,1)  {2};

  \foreach \from/\to in {n1/n2}
    \draw (\from) -- (\to);
\end{tikzpicture}}}\,%
}

\newcommand{\fourdiaone}{%
\,\mbox{\raisebox{-0.3\height}{\begin{tikzpicture}
  [scale=.07,auto=left,every node/.style={circle,scale=0.3,fill=black}]
  \node (n1) at (1,1) {1};
  \node (n2) at (9,1)  {2};
  \node (n3) at (9,9)  {2};
  \node (n4) at (1,9)  {2};

  \foreach \from/\to in {n1/n2,n2/n3,n3/n4,n4/n1}
    \draw (\from) -- (\to);
\end{tikzpicture}}}\,%
}

\newcommand{\fourdiatwo}{%
\,\mbox{\raisebox{-0.3\height}{\begin{tikzpicture}
  [scale=.07,auto=left,every node/.style={circle,scale=0.3,fill=black}]
  \node (n1) at (1,1) {1};
  \node (n2) at (9,1)  {2};
  \node (n3) at (9,9)  {2};
  \node (n4) at (1,9)  {2};

  \foreach \from/\to in {n1/n2,n2/n3,n3/n4,n4/n1,n2/n4}
    \draw (\from) -- (\to);
\end{tikzpicture}}}\,%
}
\newcommand{\fourdiathree}{%
\,\mbox{\raisebox{-0.3\height}{\begin{tikzpicture}
  [scale=.07,auto=left,every node/.style={circle,scale=0.3,fill=black}]
  \node (n1) at (1,1) {1};
  \node (n2) at (9,1)  {2};
  \node (n3) at (9,9)  {3};
  \node (n4) at (1,9)  {4};

  \foreach \from/\to in {n1/n2,n2/n3,n3/n4,n4/n1,n1/n3,n2/n4}
    \draw (\from) -- (\to);
\end{tikzpicture}}}\,%
}

\newcommand{\textdiaWWB}{%
\mbox{\raisebox{0.5ex}{\raisebox{-0.5\height}{\begin{tikzpicture}
  [scale=.04,auto=left,every node/.style={circle,scale=0.4,fill=black}]
  \node[label=below:{1},style={circle,draw,scale=1.0,fill=white}] (n1) at (1,1) {};
  \node[label=below:{2},style={circle,draw,scale=1.0,fill=white}] (n2) at (9,1)  {};
  \node[style={scale=0.6}] (n3) at (5,8)  {3};

  \foreach \from/\to in {n1/n2,n2/n3,n3/n1}
    \draw (\from) -- (\to);
\end{tikzpicture}}}}%
}
\newcommand{\inters}{\Sigma}
\newcommand{\coords}{\mathbf{R}}
\newcommand{\setofcoords}{\mathbb{V}}

\begin{document}

\title{Deriving fundamental measure theory from the virial series:\\Consistency with the zero-dimensional limit}
\author{Matthieu Marechal}
\affiliation{Institut f\"ur Theoretische Physik, Universit\"at Erlangen-N\"urnberg, Staudtstr.~7, 91058 Erlangen, Germany}
\author{Stephan Korden}
\affiliation{
Institute of Technical Thermodynamics, RWTH Aachen University, Schinkelstra\ss e~8, 52062 Aachen, Germany
}
\author{Klaus Mecke}
\affiliation{Institut f\"ur Theoretische Physik, Universit\"at Erlangen-N\"urnberg, Staudtstr.~7, 91058 Erlangen, Germany}
\date{\today}

\begin{abstract}
Fundamental measure theory (FMT) for hard particles has great potential for predicting
the phase behavior of colloidal and nanometric shapes.
The modern versions of FMT are usually derived from the
zero-dimensional limit, a system of at most one particle confined in 
a collection of cavities in the limit that all cavities shrink to the size of the particle.
In [Phys. Rev. E \textbf{85}, 041150 (2012)], a 
derivation from an approximated and resummed virial expansion was presented,
whose result was not fully consistent with the FMT from the zero-dimensional limit.
Here we improve upon this derivation and obtaining exactly the 
same FMT functional as was obtained earlier from the zero-dimensional limit.
As a result, further improvements of FMT 
based on the virial expansion can now be formulated, some of which we suggest in the outlook.
\end{abstract}

\pacs{%
 05.20.Jj,  
 82.70.Dd   
}

\maketitle

\section{Introduction}

Hard particles have been used as a reference model for molecules in most theoretical
approaches.
Moreover, recent progress~\cite{Sacanna2} in synthesis techniques have allowed the manufacture
of colloids and nano-particles of near arbitrary shape. Predicting the collective
behavior of arbitrarily shaped particles provides a challenge for theory and simulations alike,
if only because of the plethora of available shapes.

Density functional theory (DFT)~\cite{Evans} is a theory for equilibrium phase behavior of 
inhomogeneous many-particle systems. 
A DFT for mixtures of hard spheres,
fundamental measure theory (FMT),
was derived by Rosenfeld~\cite{Rosenfeld1989FMTPRL}.
More recent versions are based on a different derivation by Tarazona and Rosenfeld\cite{Tarazona1997cavities}
that demanded that the functional is exact in the so-called
zero-dimensional limit:
A collection of overlapping cavities which, as a whole, can contain at most one particle
in the limit that each cavity shrinks to the size of the particle. We will refer to the 
functional that is exact for three cavities whose intersection is nonzero as the 
0D-FMT functional.
After some modifications~\cite{Tarazona1997cavities,Tarazona2000FMT,Roth2002,Hansen-Goos2006WBII} (see also Sec.~\ref{sec:exp_and_scl}), the theory
provides a good description of
both the crystal and the high density fluid and accurately predicts the bulk freezing
transition~\cite{Oettel_Schilling_MC_FMT_HS}. 
In addition, this latest DFT
has been applied to inhomogeneous systems of hard spheres, such as 
a fluid around a hard spherical obstacle \cite{Hansen-Goos2006WBII} and the fluid-solid interface~\cite{Haertel_hs_ifce}.

Fundamental measure theory was also extended to (mixtures of) non-spherical hard particles by Rosenfeld~\cite{Rosenfeld1996oldFMT} whose description 
lacked a stable nematic phase. This artifact of FMT was repaired~\cite{Hansen-Goos2010edFMTlong,Hansen-Goos2009edFMTPRL}
by applying the Gauss-Bonnet
theorem to the intersection between two particles~\cite{[{For a similar and earlier calculation, see }]Wertheim_convex_I},
which appears in the lowest order in the excess free energy (with respect to that
of the ideal gas). The resulting functional 
has been applied to bulk phases ranging from the nematic phase for spherocylinders~\cite{Hansen-Goos2009edFMTPRL} to the crystal for (rounded) parallel hard
cubes~\cite{MM_Zimmermann_rcubes}, as well as the isotropic--nematic interface for spherocylinders~\cite{Wittmann_IN_SC} and inhomogeneous fluids of dumbbells~\cite{MM_Goetzke_db_FMT} and polyhedra~\cite{auto_MM_polyhedra}.
In addition, DFT functionals for particular shapes~\cite{Schmidt_needle_spheres,Esztermann2006}
and fixed orientations~\cite{Cuesta_paraPRL2,Martinez-Raton_para_cyl} can often be
derived more elegantly and using fewer approximations than for the general case.

Recently, an FMT functional was derived from an approximated and resummed virial expansion~\cite{Korden_deriving_FMT}.
This derivation 
justifies subsequent rescaling to obtain a better match to the next-lowest order virial diagram. A second advantage is that an accurate approximation
for the virial expansion should be valid for all external potentials, not just for the homogeneous fluid and extremely 
confined systems. Finally and most importantly, the approximation performed on the virial expansion should be amendable to
further improvements. Unfortunately, the derivation in Ref.~\cite{Korden_deriving_FMT} did not lead to the same
functional as the one from the zero-dimensional limit ~\cite{Tarazona1997cavities}, which 
is surprising as the approximated virial series is also exact~\cite{Korden_deriving_FMT} for the cavities where 0D-FMT is exact~\cite{Tarazona1997cavities}.

 In this paper, we give an improved
version of the derivation:
First, it will be clear where the combinatorial prefactors in our approximated 
virial series come from. Secondly, we improve on the calculation of the intersection of three particles surfaces,
which leads to consistency with 0D-FMT in three dimensions. 
Finally, we will consider explicitly all $d\leq3$ spatial dimensions.
The case $d<3$ deserves to be considered explicitly, since
many systems--colloidal suspensions in particular--exhibit interesting effects of reduced dimensionality~\cite{auto_Oguz_MM_confined}.
Furthermore, 
an important test for the $d=3$ functional is to evaluate it for 
the density profile of an extremely confined, quasi-$d'$-dimensional system with $d'<3$ and compare to the results from the functional 
that is obtained directly by considering $d=d'$ explicitly~\cite{Schmidt_dim_cross}.

This paper is organized as follows:
We will first introduce some notation to allow
for a reasonably compact form of our formula's in Sec.~\ref{sec:notation}. In Sec.~\ref{sec:virial}, we will 
formally define the free energy in terms of a virial expansion.
Subsequently, we will briefly describe the FMT functional \cite{Rosenfeld1989FMTPRL,Hansen-Goos2009edFMTPRL,Hansen-Goos2010edFMTlong}
in Sec.~\ref{sec:previous_derivations}. The general form of 
the functional motivates the approximation for the Mayer diagrams that we will use,
which is the main result of Ref.~\cite{Korden_deriving_FMT}. We reiterate this approximation and show its relation to 
the Ree-Hoover resummation~\cite{Ree_virial} of the Mayer diagrams in Sec.~\ref{sec:Ree-Hoover}.
In Sec.~\ref{sec:geometry}, we will use a geometric approach to derive the functional without
attempting to achieve full mathematical rigor.
In Sec.~\ref{sec:ins}, we will use the resulting geometrical expressions to resum our approximations for the Mayer diagrams
to a closed form that turns out to be exactly equal to the 0D-FMT functional derived 
by Tarazona and Rosenfeld~\cite{Tarazona1997cavities}. 
Our geometrical formulation of the functional allows us to calculate the direct correlation function 
using a formula from integral geometry, which we use to gauge the accuracy of the functional.
Subsequently, we briefly reiterate the approximations and rescalings of the 0D-FMT functional that were performed
to improve both the accuracy of FMT for certain common thermodynamic phases and the efficiency with which the
functional can be evaluated in Refs.~\cite{Tarazona1997cavities,Wertheim_convex,Hansen-Goos2010edFMTlong,MM_Goetzke_db_FMT_app_B}.
Finally, we summarize our results, argue why the current functional is successful in the light of the virial expansion 
and discuss some improvements to the functional motivated by our derivation in Sec.~\ref{sec:summary}.

\section{Notation}\label{sec:notation}

We will consider (in general) an $M$-components system of rigid particles in $1\leq d\leq 3$ dimensions.
In other words, a particle $i$ is fully characterized by its species $s_i \in \lbrace 1,\ldots,M\rbrace$,
position $\mathbf{r}_i\in\mathbb{R}^d$ and orientation, specified by a rotation $\mathcal{R}_i\in \mathbb{SO}(d)$. 
We will denote this triplet of coordinates by $\coords_i \equiv (\mathbf{r}_i,\mathcal{R}_i,s_i)$
and take $\setofcoords$ to mean the set of all accessible coordinates.

We also take the particles to have only-hard core interactions,
that is the pair-wise interaction energy $\phi(\coords_i,\coords_j)$ between particles $i$ and $j$ is such that
\[
e^{-\beta \phi(\coords_i,\coords_j)} =\left\lbrace
\begin{array}{cl}
0 & \mathcal{B}(\coords_i)\cap \mathcal{B}(\coords_j) \neq \emptyset \\
1 & \text{otherwise} \\
\end{array}\right.,
\]
where $\mathcal{B}(\coords_i)$ is the set of points in $\mathbb{R}^d$ inside a particle with coordinates $\coords_i$.
In principle, additional continuous coordinates could also be introduced and encoded in the tuple $\coords$,
for example, internal coordinates for molecules and size/shape for polydisperse systems.
We will require that the sets $\mathcal{B}(\coords)$ are convex for all $\coords$.
We will also require that the boundaries $\partial \mathcal{B}(\coords)$
are twice differentiable, such that the principal curvatures $\kappa(\mathcal{B},\mathbf{r})$ are well-defined
for all points $\mathbf{r}$ on $\partial \mathcal{B}$ (or that the boundary can be obtained from
a limiting process of twice differentiable surfaces as \emph{e.g.} in Ref.~\cite{auto_MM_polyhedra}). We will often denote $\mathcal{B}(\coords_i)$ as $\mathcal{B}_i$
and a property $\xi(\mathcal{B}_i,\mathbf{r})$ of the surface of a particle $\mathcal{B}_i$ at a point $\mathbf{r}$ by $\xi_i(\mathbf{r})$ or even $\xi_i$, if this does not
cause confusion, \emph{e.g.} the notation $\mathbf{n}_i$ for the normal vector is an abbreviation for
$\mathbf{n}(\mathcal{B}_i,\mathbf{r})$.

For any function $f$ on $\setofcoords$, we define
\[
\int_\setofcoords\! \mathrm{d}\coords\, f(\coords) \equiv \sum_s \int_{\mathbb{R}^d}\!\mathrm{d}\mathbf{r}\! \int_{\mathbb{SO}(d)}\!\mathrm{d}\mathcal{R}\, f\big((\mathbf{r},\mathcal{R},s)\big);
\]
$n$ subsequent integrals over generalized coordinates will be denoted by $\int_{\setofcoords^n} \mathrm{d}\coords^n$.
Integration over an $m$-dimensional curved hypersurface $A$ in $\mathbb{R}^d$ will be denoted by $\int_A d^m\mathbf{r}$ (with $m<d$).
The unit $d-1$ sphere will be denoted by $S_{d-1}$, and its (hyper) surface area by $\lvert S_{d-1}\rvert$,
while the spherical area of a subset $S\subset S_{d-1}$ will be denoted by $\sigma_{d-1}(S)$.

\section{Density functional theory from a virial expansion}
\label{sec:virial}

In density functional theory (DFT), the structure of the system is described by the 
density profile $\rho(\coords)$, which is defined such that the integral of $\rho$ over a subset $A\subset \setofcoords$ is the average number of
particles with generalized coordinates in $A$. Using Mayer's celebrated virial expansion, it can be shown that 
the grand-canonical free energy $\Omega$ of a system with an external potential $V_\text{ext}(\coords)$ 
can be written as a functional $\Omega[\rho_0]$ of the equilibrium density profile $\rho_0$.
This grand potential functional is defined as
\begin{equation}
\Omega[\rho]\equiv \mathcal{F}[\rho]+ \int_{\setofcoords}\!\mathrm{d}\coords\, \rho(\coords) [V_\text{ext}(\coords)-\mu_\coords] ,\label{eqn:Omega}
\end{equation}
where the chemical potential is denoted by $\mu_\coords$, which only depends on the species component $s$ of $\coords=(\mathbf{r},\mathcal{R},s)$,
see Sec.~\ref{sec:notation}. Often there will be $\coords\in \setofcoords$ where $V_\text{ext}(\coords)$ is infinite; in this case, 
we set $V_\text{ext}(\coords)\rho(\coords)=0$ if $\rho(\coords)=0$.
The intrinsic free energy $\mathcal{F}[\rho]$ consists of two parts,
\[\mathcal{F}[\rho]=\mathcal{F}_\text{id}[\rho]+\mathcal{F}_\text{exc}[\rho].
\]
The first of these, the free energy functional for an ideal gas, $\mathcal{F}_\text{id}$, reads \cite{Evans}
\[
\mathcal{F}_\text{id}[\rho]=k_B T\int_\setofcoords\! \mathrm{d}\coords\, \rho(\coords)\log[\rho(\coords)\mathcal{V}]-\rho(\coords),
\]
where $k_B$ denotes Boltzmann's constant, $T$ the temperature and $\mathcal{V}$ is the thermal volume ($1/\mathcal{V}$ 
is defined as the integral
over the momenta conjugate to $\mathbf{R}$ in the partition sum). 
Secondly, we define an 
explicit expression for the excess free energy $\mathcal{F}_\text{exc}[\rho]$ as an infinite series of Mayer diagrams~\cite{HansenMacDonald,Morita_lemmas}:
\begin{multline}
-\beta \mathcal{F}_\text{exc}[\rho]=\sum_{n=2}^\infty\, \sum_{g\in\mathcal{M}[n]} g[\rho] \\
=\twodia+ \triadia + \fourdiaone+ \fourdiatwo+ \fourdiathree + \ldots
,\label{eqn:Fexc_vir_exp}
\end{multline}
where $\beta=1/k_B T$ and the set of $n$-node Mayer diagrams $\mathcal{M}[n]$ consists of all biconnected graphs with $n$ nodes~\footnote{
Between each pair of nodes in a biconnected graph there are at least two paths of lines,
such that each node occurs at most once on each path. Equivalently, a biconnected graph
cannot be turned into a disconnected one by removing one node and all incident edges/lines.
}.
Each of these diagrams or graphs corresponds to a functional $g[\rho]$, which is constructed 
as follows:
First, label all $n$ circles or nodes of a graph $g$ with indices 1 through $n$ in some arbitrary way.
Then define the set of index pairs $P(g)\subset \lbrace 1,\ldots,n\rbrace^2$, such that $(i,j)\in P(g)$ if and only if circle $i$ and circle $j$ are connected by a line in the graph $g$.
With these definitions, the functional $g[\rho]$ reads
\begin{equation}
g[\rho]=\frac{1}{\lvert \text{Aut}(g)\rvert } \int_{\setofcoords^n}\!\mathrm{d}\coords^n\, \prod_{i=1}^n \rho(\coords_i)\!\!\! \prod_{(i,j)\in P(g)}\!\! f_\text{M}(\coords_i,\coords_j),\label{eqn:def_dia}
\end{equation}
where $f_\text{M}(\coords_i,\coords_j)=\exp[-\beta \phi(\coords_i,\coords_j)]-1$ for particles interacting with a pairwise potential $\phi$.
For the hard particles of interest, $f_\text{M}(\coords_i,\coords_j)=-1$ on overlap and $f_\text{M}(\coords_i,\coords_j)=0$ otherwise.
Finally, the group of permutations of the nodes $1,\ldots,n$ that map $g$ to a diagram with the same connectivity is denoted by $\text{Aut}(g)$ 
with order $\lvert\text{Aut}(g)\rvert$ [the number of elements in the set of $\text{Aut}(g)$]. We will also use the notation $\lvert g\rvert_L$ for the absolute value of the integral in Eqn.~(\ref{eqn:def_dia})
without the factor $1/\lvert \text{Aut}(g)\rvert$
(we use a subscript $L$, because $\lvert g\rvert_L$ is equal to the absolute
value of the diagram $g'$ formed by labeling the nodes of $g$, such that every circle is distinguishable from the others and $\text{Aut}(g')=1$).

It can be shown~\cite{Evans} that the equilibrium density profile $\rho_0$ of a system with the external potential $V_\text{ext}(\coords)$ 
is equal to the density profile that
minimizes the grand potential functional $\Omega[\rho]$ as defined in Ref.~\cite{Evans}, provided that 
the set $\mathscr{M}$ of density profiles considered in the minimization contains 
$\rho_0$ and all $\rho$ in $\mathscr{M}$ are $v$-representable.
Here, a density profile $\rho$ is called $v$-representable if
an external potential $v$ exists, such that 
$\rho$ is equal to the equilibrium density profile of the system
with external potential $v$ and the same particle--particle interactions as the system of interest.
For practical reasons, we restrict $\mathscr{M}$ to those $\rho$ for which 
the virial expansions of $\mathcal{F}_\text{exc}$ and $\frac{\delta}{\delta\rho(\coords)}\mathcal{F}_\text{exc}$ converge
(in practice, one often uses $\frac{\delta \Omega}{\delta \rho(\coords)}=0$ to perform the minimization).
This has the advantage that all $\rho$ in $\mathscr{M}$ are $v$-representable;
in fact, the external potential $v$ which represents $\rho$ is given by
$v(\coords)=\mu_\coords-\frac{\delta \mathcal{F}}{\delta \rho(\coords)}$, which is obtained from $\frac{\delta
\Omega}{\delta \rho}=0$ by rearranging.  
Unfortunately, it cannot be guaranteed that the equilibrium density profile $\rho$ is in $\mathscr{M}$, as convergence of the virial expansion 
is difficult to ascertain. However, the series in the approximation to $\mathcal{F}_\text{exc}$ we will derive below always converges 
for the equilibrium density profile. In fact, the series converges for some density profiles that should not be in $\mathscr{M}$; for example, 
the functional predicts a finite free energy for
the one-component homogeneous fluid for any packing fraction $\eta\leq 1$,
even if $\eta$ is larger than the packing fraction
in the close packed limit, $\eta_\text{cp}\leq 1$,
where the packing fraction is defined as $\eta=N v_p/V$ with
 the number of particles $N=\int_{\setofcoords}\mathrm{d}\coords\,\rho(\coords)$, the volume of a particle $v_p$ and the system volume $V$.
Of course, such a branch of (local) minima of the grand potential can be easily dismissed as unphysical
as it extends to $\eta\geq \eta_\text{cp}$.

\section{Fundamental measure theory}\label{sec:previous_derivations}

In fundamental measure theory, the excess free energy is written as a functional of the following form
\begin{equation}
\mathcal{F}_\text{exc}=\int_{\mathbb{R}^d}\!\mathrm{d}\mathbf{r}\, \Phi\big(\lbrace n_A[\rho] (\mathbf{r})\rbrace\big),\label{eqn:Fexc_wdft}
\end{equation}
where $\lbrace n_A[\rho] (\mathbf{r})\rbrace$ is a set of weighted densities
\begin{equation}
n_A[\rho](\mathbf{r})\equiv \int_\setofcoords\!\mathrm{d}\coords\, w_A(\coords,\mathbf{r}) \rho(\coords)\label{eqn:n_alpha_def}
\end{equation}
(in the remainder we will drop the argument $[\rho]$ of $n_A$).
These weight functions 
where originally derived from the low-density limit~\cite{Rosenfeld1989FMTPRL} for spheres in three dimensions,
but subsequent generalizations all have this form.
For $d=3$, the super-index $A$ takes values in $\lbrace 0,1,\ldots,3 \rbrace \cup \lbrace (\alpha,\tau,c) \,\vert\, \alpha \in \lbrace 1,2\rbrace,\ \tau\in \mathbb{N},\ c\in \lbrace 1,\ldots,d\rbrace ^\tau
\rbrace $ where $\int_{\mathbb{R}^d}\!\mathrm{d}\mathbf{r}\, w_A(\coords,\mathbf{r})$ has
dimension [length]${}^\alpha$, $\tau$ denotes the tensor rank,
and
the index $c$ denotes the tensor component. 
The weight functions $w_A(\coords,\mathbf{r})$ are distributions rather than functions.
The first, $w_d$, is defined by
\begin{equation}
\int_{\mathbb{R}^d}\!\mathrm{d}\mathbf{r}\, f(\mathbf{r}) w_d(\coords,\mathbf{r})
\equiv \int_{\mathcal{B}(\coords)}\!\mathrm{d}\mathbf{r}\, f(\mathbf{r}),
\label{eqn:def_wd_FMT}
\end{equation}
for any function $f :\mathbb{R}^d\to\mathbb{R}$,
that is, $w_d$ simply restricts the integral to the interior of the particle. 
There is also a weight function that restricts the integral to the surface of the particles,
namely, $w_{d-1}$.
\begin{equation}
\int_{\mathbb{R}^d}\!\mathrm{d}\mathbf{r}\,f(\mathbf{r}) w_{d-1}(\coords,\mathbf{r})
\equiv \int_{\partial\mathcal{B}(\coords)}\! \mathrm{d}^{d-1}\mathbf{r}\,f(\mathbf{r})
\label{eqn:def_wd-1_FMT}
\end{equation}
Similarly, $w_0$ is defined by
\begin{equation}
\int_{\mathbb{R}^d}\!\mathrm{d}\mathbf{r}\,f(\mathbf{r}) w_{0}(\coords,\mathbf{r})
\equiv \int_{\partial\mathcal{B}(\coords)}\! \frac{K(\mathbf{r})}{\lvert S_{d-1}\vert} \mathrm{d}^{d-1}\mathbf{r}\,f(\mathbf{r}),
\label{eqn:def_w0_FMT}
\end{equation}
where $K(\mathbf{r})$ is the Gaussian curvature of the surface at $\mathbf{r}$ and $\lvert S_{d-1}\vert$ the 
spherical measure of the unit $d-1$ sphere.
Finally, for $A\neq d,d-1,0$, we only give the general form of the weight functions
\[
\int_{\mathbb{R}^d}\!\mathrm{d}\mathbf{r}\, f(\mathbf{r}) w_{A}(\coords,\mathbf{r})
\equiv \int_{\partial\mathcal{B}(\coords)}\!\mathrm{d}^{d-1}\mathbf{r}\, \bar{w}_A(\coords,\mathbf{r}) f(\mathbf{r}) 
\]
where 
$\bar{w}_A(\coords,\mathbf{r})$ contain local properties of the surface at $\mathbf{r}$~\cite{Hansen-Goos2010edFMTlong}.
Clearly, the $w_A(\coords,\mathbf{r})$ for all $A$ are invariant under simultaneous translation 
of the particle coordinates $\coords$ and the position $\mathbf{r}$.

Taylor expanding the fundamental measure free energy density $\Phi(\lbrace n_\alpha\rbrace)$ around $(n_0,n_1,\cdots\big)=(0,0,\cdots)$ yields
\[
\sum_{n=0}^\infty 
\frac{1}{n!}\sum_{A_1,\ldots,A_n} \left. \frac{\partial^n \Phi}{\partial n_{A_1}\cdots \partial n_{A_n}}\right \lvert_{n_A=0}
\prod_{i=1}^n n_{A_i}.
\]
Inserting this expression into the free energy (\ref{eqn:Fexc_wdft}), we see that the free energy can be written as
\begin{align}
\mathcal{F}_\text{exc}&=\sum_{n=0}^\infty
\int_{\setofcoords^n}\! \mathrm{d}\coords^n\, \prod_{i=1}^n \rho(\coords_i) \mathcal{K}_n(\coords^n),\quad\text{where} \label{eqn:kernel_def}\\
\mathcal{K}_n(\coords^n)&=\frac{1}{n!}\sum_{A_1,\ldots,A_n}  \left. \frac{\partial^n \Phi}{\partial n_{A_1}\cdots \partial n_{A_n}}\right \lvert_{n_A=0}
\\&\times
\int_{\mathbb{R}^d}\!\mathrm{d}\mathbf{r}\, 
\prod_{i=1}^n w_{A_i}(\coords_i,\mathbf{r}),\nonumber
\end{align}
which is only nonzero if there is at least one $\mathbf{r}$ that is inside 
each particle $\mathcal{B}_i$, as a result of the range of the weight functions $w_A$. In other words, 
if we interpret (\ref{eqn:kernel_def}) as a virial expansion, then in each $n$-particle diagram
only configurations $\mathbf{R}^n$ with $\bigcap_{i=1}^n \mathcal{B}(\coords_i)\neq \emptyset$ are included.

\section{Stacks and Ree-Hoover diagrams}\label{sec:Ree-Hoover}

As we have just shown in Sec.~\ref{sec:previous_derivations}, only $n$-particle configurations $\coords^n$ with $\bigcap_{i=1}^n \mathcal{B}(\coords_i)\neq
\emptyset$ have to be taken into account in the $n$-th order virial term in order to obtain a functional of the FMT-form~(\ref{eqn:Fexc_wdft}).
For this reason, it was suggested in Ref.~\cite{Korden_deriving_FMT} to approximate a specific Mayer diagram by 
restricting the integral to configurations with a non-empty `particle stack', 
where the `stack',
introduced in Ref.~\cite{Korden_deriving_FMT}, is defined as
\[
\mathrm{St}_n(\coords^n)\equiv \bigcap_{i=1}^n \mathcal{B}(\coords_i).
\]
In Ref.~\cite{Korden_deriving_FMT}, 
only the fully connected Mayer-diagram (the diagram where each node is connected to all other nodes by Mayer-bonds)
was included in the virial expansion; all other diagrams were neglected. 
Subsequently, a somewhat involved argument was used to obtain a free energy that is of the FMT form.
Here, 
we retain all Mayer diagrams and apply the same approximation as in Ref.~\cite{Korden_deriving_FMT} to every diagram, which allows a more straightforward route using 
only the virial expansion.

The approximation to only
include $\coords^n$ with non-empty $\mathrm{St}_n(\coords^n)$ in a Mayer diagram greatly simplifies
the corresponding integrals, because it implies that each particle overlaps with each other particle or,
equivalently, $f_\text{M}(\coords_i,\coords_j)=-1$ for $1\leq i\neq j\leq n$. As a result, the product of
the Mayer bonds in each Mayer diagram $g$ with $n$ nodes, \emph{cf.} Eqn.~(\ref{eqn:def_dia}), becomes
\begin{gather}
\prod_{(i,j)\in P(g)}\!\! f_\text{M}(\coords_i,\coords_j) \to (-1)^{\lvert P(g) \rvert} \label{eqn:app_Mayer} \,
   \chi'(\coords^n)\\
\text{where}\quad\qquad \chi'(\coords^n)=   \begin{cases}
      1 & \text{if}\ \mathrm{St}_n(\coords^n) \text{is non-empty}\\
      0 & \text{otherwise}
   \end{cases}\qquad\qquad\phantom{\text{where}}\nonumber
\end{gather}
and $\lvert P(g)\rvert$ denotes the number of elements
of $P(g)$, or, equivalently, the number of lines in $g$.
In order to simplify this expression further in Sec.~\ref{sec:geometry}, we have to introduce a notion from geometry, namely the Euler characteristic.
The Euler characteristic $\chi(S)$ of a subset $S$ of $\mathbb{R}^d$ 
is equal to one when $S$ is a convex set and zero when $S$ is empty.
Therefore, $\chi'(\coords^n)$ is equal to $\chi[\bigcap_{i=1}^n \mathcal{B}(\coords_i)]$ as
we have restricted ourselves to
convex bodies, see Sec.~\ref{sec:notation},
and the intersection of any number of convex sets is either convex or empty.
As $\chi(S)$ is a topological 
invariant, $\chi(S)$ is also equal to one if $S$ is a smooth deformation of a convex set, such as the set that
we will encounter in Sec.~\ref{sec:Gauss-Bonnet}.
In other cases, $\chi(S)$ can have any
integer value  (positive, negative or zero). Therefore, $\chi'(\coords^n)=\chi[\bigcap_{i=1}^n \mathcal{B}(\coords_i)]$ does not hold in general for 
non-convex particles whose intersections can be topologically nontrivial.

Collecting all approximated Mayer diagrams with the same number of nodes, the approximation for the excess free energy from the virial expansion
(\ref{eqn:Fexc_vir_exp}) can be written as
\begin{equation}
\sum_{n=2}^\infty c_n \int_{\setofcoords^n}\!\mathrm{d}\coords^n\,  \chi\big[\textstyle\bigcap_{i=1}^n \mathcal{B}(\coords_i) \big]\, \prod_{i=1}^n
\rho(\coords_i),\label{eqn:Fexc_vir_app}
\end{equation}
where  we introduced 
 the combinatorial factor $c_n$,
\begin{equation}
c_n\equiv -\sum_{g\in \mathcal{M}[n]} \frac{1}{\lvert \text{Aut}(g)\rvert } (-1)^{\lvert P(g) \rvert}. \label{eqn:c_n_def}
\end{equation}

It remains 
to obtain a closed form for $c_n$ as a function of $n$. 
The latter problem has already been solved by Ree and Hoover in the context of their resummation of the Mayer diagrams to 
obtain an efficient algorithm for the virial expansion for homogeneous fluids of hard spheres~\cite{Ree_virial,Ree_7vir}.
They introduced a new type of diagram, which we will call a Ree-Hoover diagram, which contains 
Ree-Hoover bonds $e_\text{RH}\equiv 1-f_\text{M}$ in addition to the Mayer bonds $f_M$.
The systematic resummation of the Mayer diagrams into Ree-Hoover diagrams 
is obtained in the following way~\cite{Ree_virial}:
First, the integrand of each Mayer diagram $g$ is multiplied by a
factor $1=e_\text{RH}(\coords_i,\coords_j)-f_\text{M}(\coords_i,\coords_j)$ for every pair of nodes $(i,j)$ that is not connected by a line in $g$.
Subsequently, one expands the resulting expression in products of $e_\text{RH}(\coords_i,\coords_j)$ and $f_{M}(\coords_i,\coords_j)$ functions, and
finally, the diagrams $g$ that have the same $\lvert g\rvert_L$ are collected into one diagram, which we call
a Ree-Hoover diagram.
It can be easily seen that each Ree-Hoover diagram $g_\text{RH}$ contains the contributions of all
configurations for which a given pair of particles $i$ and $j$ is either required to overlap, 
if $i$ and $j$ are connected by an $f_\text{M}$-bond in $g_\text{RH}$, or $i$ and $j$ are not allowed to overlap, if $i$ and $j$ are 
connected by a $e_\text{RH}$-bond in $g_\text{RH}$. 
Ree and Hoover abbreviated the diagrams by leaving out the $f_\text{M}-$bonds and
denoted the $n$-particle diagram without $e_\text{RH}$-bonds by the symbol for the empty set $(\emptyset)_n$.

The Ree-Hoover resummation reduces the amount of cancellation between different diagrams in the virial series:
The configurations in a class in which $P$ is the set of overlapping pairs contribute to all Mayer diagrams $g$
for which at least each pair of nodes in $P$ is connected by a line, that is, all $g$ for which $P\subset P(g)$. Therefore,
for many classes with overlapping particles in $P$, the negative contributions to diagrams with an odd number of bonds partially
cancel the positive contributions to the Mayer diagrams with an even number of bonds.
In particular, each $n$-particle configuration that contributes to the $(\emptyset)_n$ diagram 
contributes to all Mayer diagrams, but with different prefactors (with oscillating signs), such that 
\[
(\emptyset)_n = 
\sum_{g\in \mathcal{M}[n]} \frac{1}{\lvert \text{Aut}(g)\rvert } (-1)^{\lvert P(g) \rvert} \lvert s_n\rvert_L,
\]
where $s_n$ is the fully connected \emph{Mayer} diagram. We see that the prefactor is equal to $-c_n$,
where $c_n$ is defined in Eqn.~(\ref{eqn:c_n_def}).
In other words, we would have obtained the same functional if we had started our derivation with the Ree-Hoover resummation of the Mayer diagrams,
only retained the $(\emptyset)_n$ diagrams and, finally, neglected those configurations $\coords^n$ in the integrals
of $(\emptyset)_n$ where $\mathrm{St}_n(\coords^n)=\emptyset$. Following
Tarazona and Rosenfeld~\cite{Tarazona1997cavities}, we will use the term `lost cases'
for the configurations neglected in the latter approximation.

Combinatorial techniques
for the Mayer diagrams~\cite{Ridell_Uhlenbeck}, which are beyond the scope of this work,
have been used to find a closed form for the prefactor $c_n$ as a function of $n$.
In our case, it is found~\cite{Ree_virial} that the prefactor $c_n$ is 
\begin{equation}
c_n=1/[n(n-1)]. \label{eqn:c_n_value}
\end{equation}
As an aside, we note that the prefactor of Ree-Hoover diagrams with a small number of $e_\text{RH}$-bonds can be conveniently calculated using techniques from
Refs.~\cite{Ree_Hoover_comb,Kaouche_Mayer} if the prefactor of the smallest such diagrams is known.

\section{Geometry of a stack}\label{sec:geometry}

We wish to find an expression for $\chi(\mathrm{St}_n)$ (where $\mathrm{St}_n=\textstyle\bigcap_{i=1}^n
\mathcal{B}(\coords_i)$ as before) that allows summation of the free energy (\ref{eqn:Fexc_vir_app}) to a closed form.
In order to do this we will require some concepts from integral geometry, which we will introduce along the way. 
More information and more definitions for more general (non-smooth)
convex bodies can be found in \emph{e.g.} Refs.~\cite{Schneider_Weil} and \cite{moszynska2006convex}.

As noted by Gauss for three-dimensions,
there exists a natural map, the Gauss map, from the surface of a smooth $d$-dimensional body to the $d$-dimensional unit sphere $S_{d-1}$, where 
a point $\mathbf{p}$ on the surface is mapped to the outer normal of the surface at $\mathbf{p}$. 
We will use a slight extension for non-smooth particles, where a point $\mathbf{p}$ on the surface $\partial \mathcal{B}$ of a body $\mathcal{B}$
is mapped to the `normal cone`, defined as
\[
N^+(\mathcal{B},\mathbf{p})=\lbrace \mathbf{n}\in S_{d-1} \vert \mathbf{n} \cdot (\mathbf{b}-\mathbf{p}) \leq 0\quad \forall  \mathbf{b} \in \mathcal{B} \rbrace.
\]
If $\partial \mathcal{B}$ is locally smooth at $\mathbf{p}$, $N^+(\mathcal{B},\mathbf{p})$ is a set with one element, the unique normal at $\mathbf{p}$.
Note, that the definition of the normal cone $N^+(\mathcal{B},\mathbf{p})$ is only useful for convex particles: $N^+(\mathcal{B}_{n{c}},\mathbf{p})=\emptyset$ for a point
$\mathbf{p}$ on the concave part of the surface of a non-convex body $\mathcal{B}_{nc}$.
We can straightforwardly extend this definition to (Borel) subsets $A$ of $\mathbb{R}^d$ to obtain the normal cone of $A$,
\begin{equation}
N^+(\mathcal{B},A)=\bigcup_{\mathbf{p} \in A\cap\partial \mathcal{B} } N^+(\mathcal{B},\mathbf{p}). \label{eqn:def_normal_cone}
\end{equation}
The spherical measure $\sigma_{d-1}(\cdot)$ of the normal cone relative to the total measure of the $(d-1)$ sphere $\lvert S_{d-1}\rvert$ will play a central role in our derivation, so
we will use a separate symbol $\gamma(\mathcal{B},\cdot)$ for the corresponding measure, 
\[
\gamma(\mathcal{B},A)\equiv
\frac{\sigma_{d-1}\big(N^+(\mathcal{B},A)\big)}{\lvert S_{d-1}\rvert}.
\]
for any (Borel) subset $A\subset \mathbb{R}^d$.
Applying this measure to the stack, we can write 
\begin{equation}
\chi\big({\textstyle\bigcap_{i=1}^n \mathcal{B}_i}\big)=
\gamma(\mathrm{St}_n,\mathbb{R}^d)=
\gamma(\mathrm{St}_n,\partial\mathrm{St}_n).\label{eqn:gamma_eq_chi}
\end{equation}
Here and in the remainder of this section (Sec.~\ref{sec:geometry}), 
we write $\mathcal{B}_i$ instead of $\mathcal{B}(\coords_i)$.
Of course, if $\mathrm{St}_n=\emptyset$, $N^+(\mathrm{St}_n,\mathbf{p})=\emptyset$ and $\gamma(\mathrm{St}_n,\mathbf{p})=0$ for all $\mathbf{p}\in\mathbb{R}^d$,
while $N^+(\mathrm{St}_n,\mathbb{R}^d)$ is the full unit sphere if $\mathrm{St}_n$ is non-empty, such that $\gamma(\mathrm{St}_n,\mathbb{R}^d)=1$,
which proves Eqn.~(\ref{eqn:gamma_eq_chi}) for a stack of \emph{convex} particles.

\begin{figure}
\includegraphics[width=0.45\textwidth]{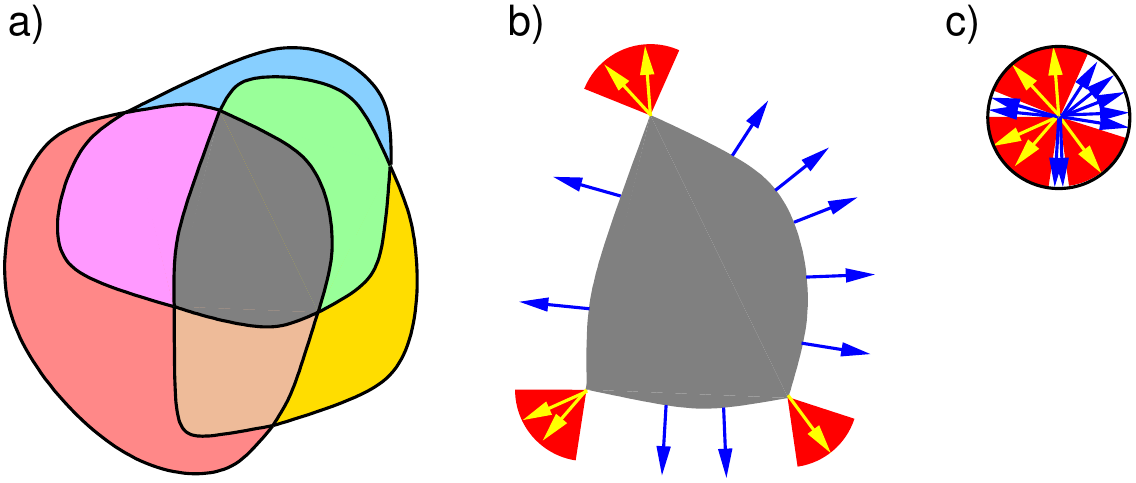}
\caption{ 
 (a) Example of a `stack' in two dimensions: The dark area denotes the `stack' of three arbitrary convex bodies, defined as the intersection between the
bodies. The main approximation necessary to derive FMT is to restrict the multi-dimensional integrals in the Mayer diagrams 
to include only those configurations as shown here, that is, where the intersection between all bodies is non-empty.
(b) and (c) The union of all normal vectors of the surface of a convex body (b) is just the $(d-1)$-dimensional unit hyper-sphere (c),
(a circle with unit radius for $d=2$). The set of vectors normal to a point $\mathbf{p}$ on the surface of a stack, $N^+(\mathbf{p})$, can either contain
a single normal vector for a point $\mathbf{p}$ on one of the surfaces of the particles that constitute the stack (examples are 
the dark arrows) or a $(k-1)$-dimensional set of normal vectors for $\mathbf{p}$ on the intersection between $k$ surfaces
(\emph{e.g.} the sectors containing the light arrows for $k=2$).
\label{fig:decomp}}
\end{figure}

In our case, this is useful because of the following decomposition:
\begin{multline}
N^+(\mathrm{St}_n,\partial\mathrm{St}_n)
\\
\left. \begin{aligned}
&=
\bigcup_{i=1}^n N^+\big(\mathrm{St}_n,\partial\mathcal{B}_i \cap \mathrm{St}^\cup_{n\setminus i}\big) 
\\&\cup
\bigcup_{\substack{i,j=1\\i< j}}^n N^+\big(\mathrm{St}_n,\partial\mathcal{B}_i \cap \partial \mathcal{B}_j\cap \mathrm{St}^\cup_{n\setminus i,j}\big) 
\\&\cup
\ldots 
\\&\cup 
\bigcup_{\substack{i_1,\ldots,i_m=1\\i_1<\ldots < i_m}}^n N^+\big(\mathrm{St}_n,\partial\mathcal{B}_{i_1} \cap\ldots \cap \partial \mathcal{B}_{i_m}\cap
\mathrm{St}^\cup_{n\setminus i_1,\ldots,i_m}\big) 
\end{aligned}\right.
\hspace*{-1.7em}\label{eqn:union}
\end{multline}
where we have defined the \emph{open} subset of $\mathbb{R}^d$,
\[
\mathrm{St}^\cup_{n\setminus i_1,\ldots,i_m}\equiv \bigcap_{\substack{\nu=1\\\nu\neq i_1, \ldots, \nu\neq i_n}}^n \text{int}(\mathcal{B}_\nu).
\]
Here, 
$\text{int}(\mathcal{B})$ is the interior of the body $\mathcal{B}$ \emph{i.e.} $\text{int}(\mathcal{B})=\mathcal{B}\setminus \partial \mathcal{B}$.
An example of the decomposition (\ref{eqn:union}) in $d=2$ dimensions is shown in Fig.~\ref{fig:decomp}.

In the remainder of this section, we will exclude some pathological $n$-particle configurations with a vanishing 
contribution to the free energy, namely those for which one or more surfaces $\partial \mathcal B_i$
are tangent to either another surface $\partial \mathcal B_j$ or the intersection between two or more of the other surfaces or 
for which the intersection between $d$ particles in $d$ dimensions lies on one of the surfaces.
With this restriction, the intersection between $k$ surfaces is always a $(d-k)$-dimensional subset of $\mathcal{R}^d$.
As a result, the number of elements in the union (\ref{eqn:union}), $m$, is at most $\min\lbrace n,d\rbrace$ in $d$ dimensions~%
\footnote{
The number $m$ can be less than {$\min\lbrace n,d\rbrace$}, if one or more of the bodies lie inside other bodies in the stack.
}.
Note, that the surface of the stack is non-smooth near all points $\mathbf{p}$ on the intersections of $k$ surfaces $\partial B_i$
(for $k\geq 2$),
and that $N^+(\mathrm{St}_n,\mathbf{p})$ is a $(k-1)$-dimensional set for such points $\mathbf{p}$.
For \emph{convex} particles, we know that every unit vector is an outer normal vector to the surface in exactly one point, which implies that the sets in the
union (\ref{eqn:union}) are pairwise disjoint
(more precisely $\sigma_{d-1}(A \cap B)=0$ for any two sets $A$, $B$ in the union (\ref{eqn:union}) with $A\neq B$). Therefore,
\begin{multline*}
\gamma(\mathrm{St}_n,\partial\mathrm{St}_n) =\\ 
\sum_{k=1}^m
\sum_{\substack{i_1,\ldots,i_k=1\\i_1<\ldots < i_k}}^n \gamma\big(\mathrm{St}_n,\partial\mathcal{B}_{i_1} \cap\ldots \cap \partial \mathcal{B}_{i_k}\cap
\mathrm{St}^\cup_{n\setminus i_1,\ldots,i_k}\big),
\end{multline*}
which is also obvious from the fact that $\gamma(B,\cdot)$ is additive~\cite{moszynska2006convex}.
The decomposition of $\gamma$ is a special case of the decomposition of local Minkowski functionals~\cite{[{See \emph{e.g.} }]Mecke_moments_boolean}.
Consider a  $d-k$ dimensional intersection $\inters_{d-k}$ of
$k$ surfaces of some convex bodies $\lbrace \mathcal{B}_{j}\rbrace_{j=1}^k$ for $2\leq k\leq d$.
Using the Lebesgue integral 
with $\gamma(B,\cdot)$ as the integration measure,
we can now formally define a generalized weight function $w^{[k]}(\mathcal{B}_{1}, \dots, \mathcal{B}_{k},\mathbf{r})$ 
by
\begin{multline}
\int_{\mathbb{R}^d}\!\mathrm{d}\mathbf{r}\, f(\mathbf{r}) w^{[k]}(\mathcal{B}_{1},\dots,\mathcal{B}_{k};\mathbf{r})
\\\equiv 
\frac{1}{k!}\int_{\inters_{d-k}\subset \mathbb{R}^d}\!\gamma(\mathcal{B}_{1}\cap\dots\cap\mathcal{B}_{k},\mathrm{d}\mathbf{p})\, f(\mathbf{p}) 
\label{eqn:def_w_k_gen}
\end{multline}
for any piecewise continuous function $f(\mathbf{r})$ on $\mathbb{R}^d$.
We have absorbed a factor $1/k!$ into the generalized weight function
that would have otherwise appeared explicitly in the excess free energy functional $\mathcal{F}_\text{exc}$ in Sec.~\ref{sec:ins}. 

With this definition (\ref{eqn:def_w_k_gen}), we can write the normalized normal cone area of the stack as
\begin{multline}
\gamma\big(\mathrm{St}_n,\partial\mathrm{St}_n\big)=\\
\int_{\mathbb{R}^d}\!\!  \mathrm{d}\mathbf{r}
\sum_{k=1}^m
\sum_{\substack{i_1,\ldots,i_k=1\\i_1\neq\ldots \neq i_k}}^n \!\!\!
w^{[k]}\big(\mathcal{B}_{i_1},\!\cdots\!,\mathcal{B}_{i_k};\mathbf{r}\big)
\!\!\!\!\!\!\!\prod_{\substack{j=1\\j\notin \lbrace i_1,\cdots,i_k\rbrace}}^n\!\!\!\!\!\!\! w_d(\mathcal{B}_{j},\mathbf{r}),\label{eqn:normal_cone_w}
\end{multline}
with the volume weight function $w_d$ from Eqn.~(\ref{eqn:def_wd_FMT}).
This expression suffices to resum the approximated Mayer diagrams in a closed form to obtain a functional. However,
the expressions for the $k$-weight functions (\ref{eqn:def_w_k_gen}) for $k\geq 2$ are not very useful for explicit calculations
and they are difficult to compare to the FMT weight functions. Therefore, we specialize to cases where we can give explicit expressions in
Sect.~\ref{sec:special_cases}.

\subsection{Special cases}
\label{sec:special_cases}

In this section, we will restrict ourselves to the relevant cases
for the spatial dimensions $1\leq d\leq 3$, namely $k=d=1$; $k=d$ for $d=2,3$; $k=1$ for general $d>1$ and, finally, $k=2$ for $d=3$.
For these cases, we will obtain explicit expressions for
the resulting weight functions, which
are summarized in Table~\ref{tbl:kbody_w}. 
The general case will be discussed elsewhere~\cite{arXiv_Korden_DFT_N_dims}.

\subsubsection{One dimension}
\label{sec:one_D}

\begin{figure}
\includegraphics[width=0.99\columnwidth]{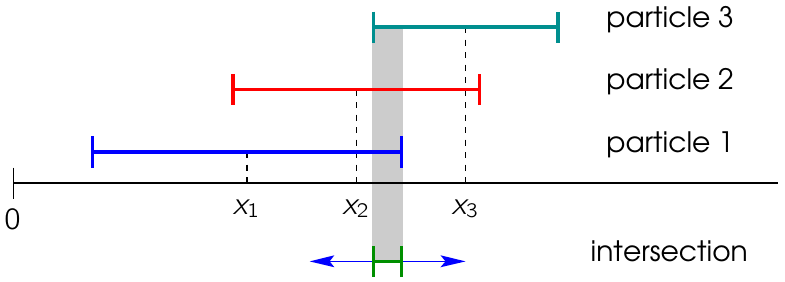}
\caption{
In one dimension, `particles' are line intervals and only two normals are possible,
$+1$ and $-1$, as shown by the arrows. The `stack' of the three particles shown
is the line interval labeled `intersection'.
\label{fig:intersection_1D}}
\end{figure}

For $d=1$, every $\mathcal{B}_i=[a_i,b_i]$ for some real numbers $a_i$ and $b_i$ and we have $m=1$ since the boundaries of the particles cannot
intersect.
The stack for $d=1$ is illustrated in Fig.~\ref{fig:intersection_1D}.
It is  non-empty if $\min_i b_i>\max_i a_i$,
in which case 
 the normal cone of the stack [as defined in Eqn.~(\ref{eqn:def_normal_cone})] 
consist of the normal vectors $+1$ at $\min_i b_i$ and $-1$ at $\max_i a_i$.
(The $0$-dimensional unit sphere consist of the two points $\pm 1$ and $\lvert S_0\rvert =2$.) 
The normal cone area for $\delta B_i$ in $d=1$ dimensions has the form (\ref{eqn:normal_cone_w}) with
\[
w^{[1]}(\mathcal{B}_{i};{x})
\\\equiv 
\sum_{{p}\in \lbrace a_{i},b_{i}\rbrace}
\tfrac{1}{2}\delta({x}-{p}).
\]

\subsubsection{Intersection between $d$ particles ($k=d$)}

We will first consider $k=d$ intersecting surfaces for general $d$.
The set $\inters_{d-k}=\inters_0$
is then a set of discrete points at which the boundaries intersect.
Therefore, we can immediately write down the $w^{[k]}$,
\begin{equation}
w^{[d]}(\mathcal{B}_{i_1},\dots,\mathcal{B}_{i_k};\mathbf{r})
\\\equiv 
\sum_{\mathbf{p}\in \inters_0}
\frac{\sigma_{d-1}\big(N^+(\mathrm{St}_n,\mathbf{p})\big)}{|S_{d-1}|k!}\delta(\mathbf{r}-\mathbf{p}),
\label{eqn:def_wd_FMMT}
\end{equation}
where explicit expressions for $\sigma_{d-1}\big(N^+(\mathrm{St}_n,\mathbf{p})\big)$ are given below for $d=2,3$.
Note, that $w^{[1]}$ for $d=1$ is consistent with the expression obtained in Sec.~\ref{sec:one_D}.

\subsubsection{No intersection ($k=1$)}

\begin{figure}[b!]
\hspace*{\stretch{1}}\includegraphics[width=0.45\textwidth]{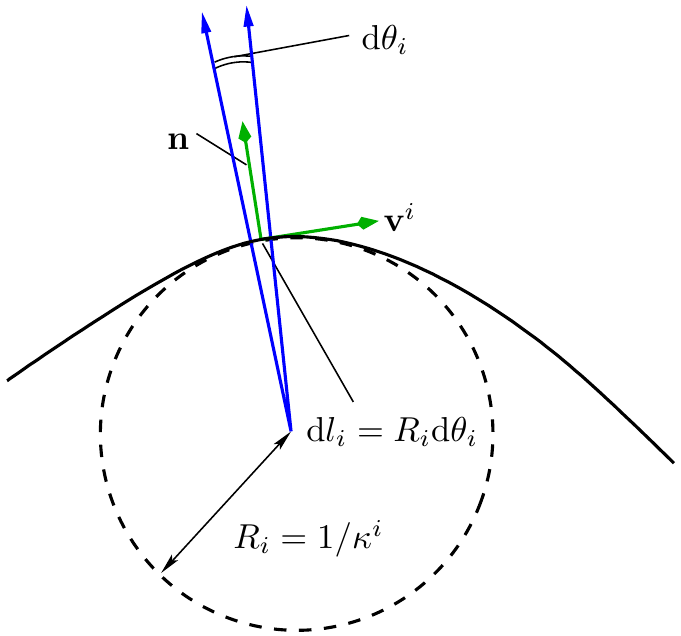}\hspace*{\stretch{1}}

\caption{ Illustration of the equality of $\gamma$ and $\Phi_0$, see Eqn.~(\ref{eqn:Phi0def}). 
The radius of curvature $R_i=1/\kappa^i$ in principal direction $\mathbf{v}^i$ at some point $\mathbf{p}$
is illustrated by the dashed circle.
The intersection between the surface $\partial \mathcal{B}$ of a particle
$\mathcal{B}$ and the plane spanned by $\mathbf{v}^i$ and the normal vector $\mathbf{n}$ at $\mathbf{p}$ is 
denoted by the black curve. The extension in direction $\mathbf{v}_i$ of the normal cone to an infinitesimally small patch $\mathrm{d}A$ of the surface near $\mathbf{p}$ is denoted by
the angle $\mathrm{d}\theta_i\simeq \mathrm{d}l_i/R_i=\kappa^i \mathrm{d}l_i$,
where $\mathrm{d}l_i$ is the extension of the patch in direction $\mathbf{v}^i$.
Now, the spherical area of the normal cone to $\mathrm{d}A$ can be seen to be equal to the Gaussian curvature integrated over $\mathrm{d}A$:
$\lvert S_{d-1}\rvert \gamma(\mathcal{B},\mathrm{d}A)\simeq \prod_i \mathrm{d}\theta_i\simeq
\prod_i\kappa_i\mathrm{d}l_i\simeq\Phi_0(\mathcal{B},\mathrm{d}A)\lvert S_{d-1}\rvert $.
\label{fig:Gauss_map}
}
\end{figure}

In order to discuss $k=1$ for $d>1$ [the case $k=d=1$ is already covered by Eqn.~(\ref{eqn:def_wd_FMMT})],
 we will use the equality (\cite{moszynska2006convex}, page 608 in \cite{Schneider_Weil}) of $\gamma$ and $\Phi_0$,
one of the curvature measures~\cite{Schneider_curvature},
which for a smooth body and some (Borel) subset $A\subset \mathbb{R}^d$ reads
\begin{equation}
\Phi_0(\mathcal{B},A)=\int_{\partial B\cap A}\! \mathrm{d}^{d-1}\mathbf{r}\, \frac{K(\mathbf{r})}{\lvert S_{d-1}\rvert} = \gamma(\mathcal{B},A)
\label{eqn:Phi0def}
\end{equation}
where $K(\mathbf{r})$ is the Gaussian curvature at $\mathbf{r}$: $K \equiv \prod_{j=1}^{d-1} \kappa^j$,
where the principal curvature in direction $\mathbf{v}^j$ is denoted by $\kappa^j$ for $1\leq j\leq d-1$.
The equality between the integrated Gaussian curvature $\Phi_0(\mathcal{B},A)$ and the normal cone area $\gamma(\mathcal{B},A)$ (which,
in its original version in $d=3$ dimensions is due to Gauss) is illustrated
in Fig.~\ref{fig:Gauss_map}.
Using this equality,
we can define $w^{[1]}$ as
\begin{equation}
w^{[1]}(\mathcal{B}_i,\mathbf{r})=\frac{K_i(\mathbf{r})}{\lvert S_{d-1}\rvert} w_{d-1}(\mathcal{B}_i,\mathbf{r}),\label{eqn:def_w_k_eq_1}
\end{equation}
which is equal to the FMT weight function $w_0$, see Eqn.~(\ref{eqn:def_w0_FMT}).
Here, we used the weight function $w_{d-1}$ as defined in Eqn.~(\ref{eqn:def_wd-1_FMT}).

\subsubsection{Two dimensions}

For the case $k=d=2$,
the normal cone of the intersection point is just the arc spanned by the normal vectors $\mathbf{n}_1$ and $\mathbf{n}_2$
of the two intersecting surfaces and
the arc length reads
\begin{equation}
\sigma_{1}\big(N^+(\mathrm{St}_n,\mathbf{p})\big)=\arccos(\mathbf{n}_1\cdot\mathbf{n}_2).
\end{equation}
Using this arc length in Eqn.~(\ref{eqn:def_wd_FMMT}) for $d=2$, the weight function $w^{[2]}$ is defined.
For the remaining case, $k=1$ in $d=2$ dimensions, we can use the expression (\ref{eqn:def_w_k_eq_1}) for $w^{[1]}_0$ given above.

\subsubsection{Three dimensions: the Gauss-Bonnet theorem}
\label{sec:Gauss-Bonnet}

\begin{figure}
\includegraphics[width=0.99\columnwidth]{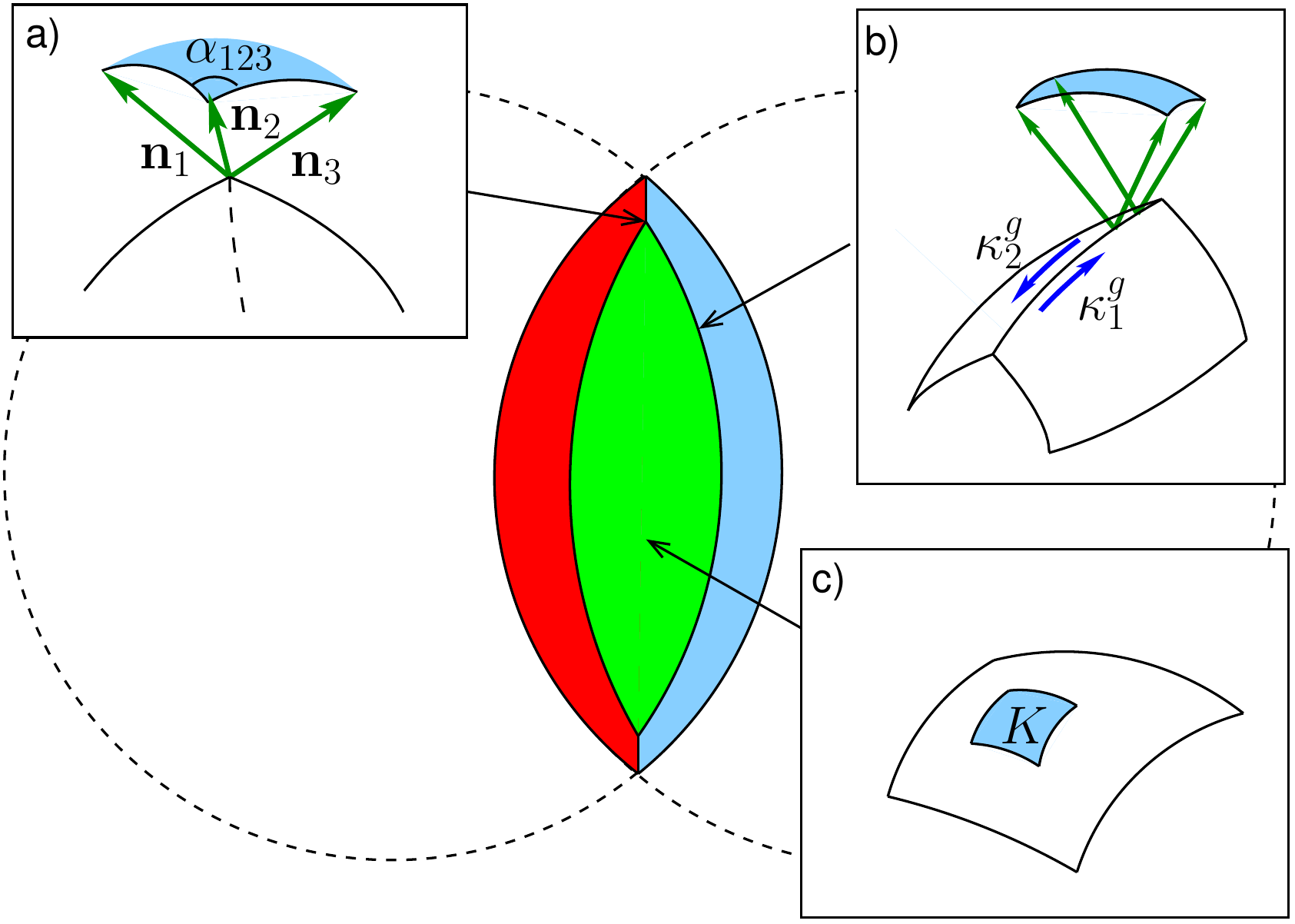}
\caption{The intersection between three particles (spheres in this case) in
three dimensions (the surfaces of the three spheres have different colors for clarity). For $d=3$, there are three contributions that are
distinguished by the number of intersecting surfaces $k$ as shown in the 
three panels: (a) $k=3$ (b) $k=2$ and (c) $k=1$. 
For $k=3$ surfaces that intersect at a point $\mathbf{p}$, the 
normal cone $N^+(\mathbf{p})$ is a spherical triangle spanned by the 
normals of the intersecting surfaces at $\mathbf{p}$ depicted by the arrows in (a). For $k=2$
and $k=1$, the contribution of a small section of the intersection (b) or the surfaces itself (c)
is indicated by the filled area. In (b), we also indicated the geodesic curvatures $\kappa^g_1$ and $\kappa^g_2$
of the
paths indicated by arrows with the corresponding labels.
\label{fig:intersection_3D}}
\end{figure}

The intersection of three bodies in three dimensions
with the three different types of contributions ($k=1,2,3$) to the normal cone
is shown in Fig.~\ref{fig:intersection_3D}.

First,  the weight function $w^{[1]}$ as defined in (\ref{eqn:def_w_k_eq_1}) can be used.
Secondly, the normal cone for $d=k=3$ 
is a spherical triangle spanned by the normal vectors
$\mathbf{n}_1$, $\mathbf{n}_2$ and $\mathbf{n}_3$ of the three intersection surfaces. 
Its area reads
\[
\sigma_{2}\big(N^+(\mathrm{St}_n,\mathbf{p})\big)=\alpha_{123}+\alpha_{231}+\alpha_{312}-\pi  \quad\text{for $d=3$},
\]
which can be inserted in Eqn.~(\ref{eqn:def_wd_FMMT}) for $d=3$ to define the weight function $w^{[3]}$.
Here, $\alpha_{ijk}$ denotes the dihedral angle, as depicted in Fig.~\ref{fig:intersection_3D}(a), which reads
\[
\alpha_{ijk}=\arccos (\mathcal{N}[\mathbf{n}_i\times\mathbf{n}_j]\cdot\mathcal{N}[\mathbf{n}_k\times\mathbf{n}_j]),
\]
where $\mathcal{N}[\mathbf{v}]$ for $\mathbf{v}\in \mathbb{R}^d$ denotes the normalized vector $\mathbf{v}/\lvert \mathbf{v}\rvert$
and the cross product (in $d=3$ only) is denoted by $\times$.

Finally, the normal cones for $k=2$ in three dimensions, or $1<k<d$ for general $d\geq 3$, contain continuous $(d-k)$-dimensional sets of points that contribute to the total normal cone,
which requires a different way of calculating the total normal cone area of the stack. One option would be to use Chern's direct approach~\cite{Chern_kinematic}.
Instead, we will perform a less involved calculation using the Gauss-Bonnet theorem, which reads (in three dimensions),
\begin{equation}
\int_{S}  K \mathrm{d}A + \sum_k\int_{\partial S_k} \kappa_g \mathrm{d} l + \sum_n \angle_{n}  = 2 \pi \chi(S),\label{eqn:GaussBonnet_gen}
\end{equation}
where $S$ is a compact twice differentiable two-dimensional surface %
 bounded by an oriented curve $\partial S$ consisting of $M$ smooth sections  $\partial S_k$, while the curve turns by
$\angle_{n}$ at the intersection between sections $\partial S_{n-1}$ and $\partial S_n$ (and $\partial S_0 \equiv \partial S_M$). Furthermore, $K$ is
the Gaussian curvature and $\kappa_g$ is the geodesic curvature 
on the smooth sections of $\partial S$. Finally, $\chi(S)$ is the Euler characteristic of the surface. 

The Gauss-Bonnet theorem cannot be directly applied to the stack because it is not smooth (the Gaussian curvature is not well-defined
on the intersection between two or three surfaces),
so we will consider the `tube' or the surface parallel to a subset $A$ of the surface of the stack:
\[
t_\epsilon(A)=\bigcup_{\mathbf{p}\in A} \epsilon N^+(\mathrm{St}_n,\mathbf{p})+\mathbf{p}
\]
(where scaling a set by a constant and summing a set and a vector implies performing the operation
to each element separately: $\mathbf{p}+\epsilon C:=\epsilon C+\mathbf{p}:=\lbrace \mathbf{p}+\epsilon\mathbf{r} | \mathbf{r}\in C\rbrace$
for any set $C\subset \mathbb{R}^d$ ).
The parallel body is defined by 
\begin{equation}
(\mathcal{B})_\epsilon=\lbrace \mathbf{p}+\mathbf{r} : \mathbf{p}\in \mathcal{B},\,\mathbf{r}\in \mathbb{R}^d,\,\lvert\mathbf{r}\rvert \leq \epsilon \rbrace.
\label{eqn:parallel_body}
\end{equation}
Note 
that the normal cone of a subsection $S$ of $\partial \mathrm{St}_n$ is the same as 
the normal cone of the surface parallel to $S$: $N^+\big(t_\epsilon
S,(\mathrm{St}_n)_\epsilon\big)=N^+(S,\mathrm{St}_n)$; however, $t_\epsilon S$ is always a twice-differentiable $(d-1)$-dimensional
hypersurface even if $S$ is not.
With these definitions, we can consider the remaining case: $k=2$ intersecting surfaces in $d=3$ dimensions.
A very similar calculation was also used to decompose the Mayer bond
into weight functions~\cite{Wertheim_convex,Hansen-Goos2010edFMTlong}.

We will consider a subsection $\delta C$ of the curve $\inters_2=\partial B_{i_1}\cap \partial B_{i_2}$
and consider the normal cone of $t_\epsilon \delta C$ [the latter is denoted by the light blue area in Fig.~\ref{fig:intersection_3D} b)].
The direct approach, integrating the Gaussian curvature $K$ over $t_\epsilon \delta C$, has been 
followed in Refs.~\cite{Wertheim_convex_I,Korden_deriving_FMT} and will not be repeated here.

As a second approach~\cite{Wertheim_convex_I}, which connects to the deconvolution of the Mayer bond in Ref.~\cite{Hansen-Goos2010edFMTlong}, the Gauss-Bonnet
theorem, Eqn.~(\ref{eqn:GaussBonnet_gen}) can be applied for $S=t_\epsilon \delta C$. The boundary $\partial S$ consists of the arcs
\begin{align*}
\partial S_1&=\lbrace \mathbf{a}+\epsilon[\cos\phi\mathbf{e}_1(\mathbf{a})+\sin\phi\mathbf{e}_2(\mathbf{a})] : \lvert \phi\rvert <
\phi_{ij}(\mathbf{a})/2 \rbrace,\\
\partial S_3&=\lbrace \mathbf{b}+\epsilon[\cos\phi\mathbf{e}_1(\mathbf{b})+\sin\phi\mathbf{e}_2(\mathbf{b})] : \lvert \phi\rvert <
\phi_{ij}(\mathbf{b})/2 \rbrace,
\end{align*}
 which have zero geodesic curvature,
and the curves
\begin{align*}
\partial S_2 &=\lbrace \mathbf{r}+\epsilon\mathbf{n}_i(\mathbf{r}) \vert \mathbf{r}\in \delta C\rbrace\quad \text{and}\\
\partial S_4 &=\lbrace \mathbf{r}+\epsilon\mathbf{n}_j(\mathbf{r}) \vert \mathbf{r}\in \delta C\rbrace,
\end{align*}
whose geodesic curvatures
reduce to those of $\delta C$ on the two respective surfaces when $\epsilon\to0$.
Here, $\mathbf{a}$ and $\mathbf{b}$ are the end points of the curve $\delta C$; the angle $\phi_{ij}$ is defined as $\phi_{ij}\equiv\arccos(\mathbf{n}_i\cdot\mathbf{n}_i)$; $\mathbf{e}_1=(\mathbf{n}_1+\mathbf{n}_2)/\sqrt{2(1+\mathbf{n}_1\cdot\mathbf{n}_2)}$ 
and $\mathbf{e}_2=(\mathbf{n}_1-\mathbf{n}_2)/\sqrt{2(1-\mathbf{n}_1\cdot\mathbf{n}_2)}$.
It can easily be seen that the angles $\angle_n$ between the curves $\partial S_n$ and $\partial S_{n+1}$ are 
equal to $\pi/2$ for all $n$. Applying the Gauss-Bonnet theorem to $S=t_\epsilon \delta C$, rearranging
and inserting the above expressions for the geodesic curvatures and angles $\angle_n$, we obtain
\begin{multline*}
\lim_{\epsilon\to 0} \int_{S}\!\mathrm{d}^2\mathbf{r}\, K(\mathbf{r}) =\\
 \lim_{\epsilon\to 0} \Big[2\pi-4\times\pi/2 - 
\int_{\partial S_2} \!\mathrm{d}^1\mathbf{r}\,\kappa^g_i(\mathbf{r}) - \int_{\partial S_4} \!\mathrm{d}^1\mathbf{r}\,\kappa^g_j(\mathbf{r})
 \Big]\\=\int_{\delta C}\! \mathrm{d}^1\mathbf{r}\, [\kappa^g_i(\mathbf{r})+\kappa^g_j(\mathbf{r})].
\end{multline*}
Therefore, the generalized weight function for $k=2$ and $d=3$ is defined by
\[
\int_{\mathbb{R}^d}\!\mathrm{d}\mathbf{r}\,f(\mathbf{r}) w^{[2]}(\mathcal{B}_i,\mathcal{B}_j,\mathbf{r})=
\int_{\partial \mathcal{B}_i\cap\partial \mathcal{B}_j}\!\mathrm{d}^1\mathbf{r}\,
\frac{\kappa^{g}_{i}+\kappa^g_j}{8\pi}\,f(\mathbf{r})
\]
This can be seen to be equal to the result from Ref.~\cite{Korden_deriving_FMT} using the explicit expression for the geodesic curvature 
from Refs.~\cite{Wertheim_convex_I,Hansen-Goos2010edFMTlong}, 
\[
\kappa^{g}_{i}+\kappa^g_j=
\frac{\kappa_i^{I\!I}(\mathbf{v}^I_i\cdot\mathbf{n}_j)^2 +\kappa_i^I(\mathbf{v}^{I\!I}_i\cdot\mathbf{n}_j)^2}{\lvert \mathbf{n}_i\times\mathbf{n}_j \rvert\,(1+\mathbf{n}_i\cdot\mathbf{n}_j)} + (i\leftrightarrow j).
\]
where $(i\leftrightarrow j)$ means the preceding expression with $i$ and $j$ interchanged. This concludes the calculation of the $k$-body weight function
in the form given in Eqn.~(\ref{eqn:def_w_k_gen}) for the special cases $k=1,d$ for any $d$ and $1\leq k\leq d$ for $1\leq d\leq 3$.
The weight function  for $k=2$ and $d=3$ was
further simplified
in Refs.~\cite{Wertheim_convex,Hansen-Goos2010edFMTlong} as we will show in the Sec.~\ref{sec:simpl} for general $k$ and $d$.

\begin{table}

{\setlength{\arraycolsep}{0pt}
\newlength{\bla}\setlength{\bla}{0.5em}
\def\blaspace{\hspace*{1.7em}}
\noindent\begin{tabular}{|@{}>{\blaspace}c<{\rule{0pt}{2.5em}}@{}>{\blaspace}c<{}>{\blaspace}c<{\blaspace}@{\extracolsep{\fill}}|}
\hline
$d$ &
$k$ &
${Q}^{[k]}_d(\mathcal B_1,\dots,\mathcal{B}_k;\mathbf{r})$
\\[1.4em]
\hline
1 & 1 &
$\displaystyle\frac{1}{2}$\\[1.4em]
\hline
\raisebox{-0.0em}{2} & 1 &
$\displaystyle\frac{K_1}{2\pi}$ \\[1.0em]
 & 2 &
$\displaystyle \sqrt{1-(\mathbf{n}_1\cdot\mathbf{n}_2)^2}\;
\frac{\arccos(\mathbf{n}_1\cdot\mathbf{n}_2)}{4\pi}$ \\[1.4em]
\hline
3 & 1 & 
$\displaystyle
\frac{K_1}{4\pi}
$ \\[1em]
 & 2 & 
$\displaystyle
\lvert \mathbf{n}_1\times\mathbf{n}_2\rvert\,
\frac{\kappa^{g}_{1}+\kappa^g_2}{8\pi}
 $ \\[1em]
 & 3 & 
$\displaystyle
\lvert\mathbf{n}_1\cdot (\mathbf{n}_2\times\mathbf{n}_3)\rvert\,
\frac{\alpha_{123}+\alpha_{231}+\alpha_{312}-\pi}{24\pi}
$ \\[1.4em]
\hline
\end{tabular}}

\caption{
As shown in the text, all $k$-body weight functions, $w^{[k]}(\mathcal B_1,\dots,\mathcal{B}_k;\mathbf{r})$ have the form
$Q^{[k]}_d(\mathcal B_1,\dots,\mathcal{B}_k;\mathbf{r})\prod_{i=1}^k w_{d-1}(\mathcal{B}_i,\mathbf{r})$
in $d$ spatial dimensions
with the factors $Q^{[k]}_d$ that are listed in this table.
See text for the definitions of the symbols; $K_1$, $\mathbf{n}_i$, $\kappa_i^g$ and $\alpha_{ijk}$ 
depend implicitly on the point $\mathbf{r}$ and the body/bodies with the corresponding index/indices.
For comparison to Refs.~\cite{Hansen-Goos2010edFMTlong,Wertheim_convex,Tarazona1997cavities,
MM_Goetzke_db_FMT_app_B},
we wrote $[\mathbf{n}_1]=1$ and $[\mathbf{n}_1,\mathbf{n}_2]=\sqrt{1-(\mathbf{n}_1\cdot\mathbf{n}_2)^2}$, which are valid for general $d$, as well as 
$[\mathbf{n}_1,\mathbf{n}_2]=\lvert \mathbf{n}_1\times \mathbf{n}_2 \rvert$ and
$[\mathbf{n}_1,\mathbf{n}_2,\mathbf{n}_3]=\lvert \mathbf{n}_1\cdot(\mathbf{n}_2\times \mathbf{n}_2)\rvert$
which are valid only for $d=3$.
\label{tbl:kbody_w}}
\end{table}

\subsection{Simplifying the integral over the intersection}
\label{sec:simpl}

The definition of the $k$-body weight function contains a cumbersome integral
over the intersection of $k$ surfaces.
This integral can be simplified using
\begin{multline}
\int_{\partial B_1\cap\dots\cap\partial B_k}f(\mathbf{r})\mathrm{d}^{d-k}\mathbf{r}\\
=\int_{\mathbb{R}^d} [\mathbf{n}_1,\dots,\mathbf{n}_k] f(\mathbf{r})\prod_{i=1}^k w_{d-1}(\mathcal B_i,\mathbf{r})\mathrm{d}\mathbf{r}
\label{eqn:simpl_intersect}
\end{multline}
for any function $f:\mathbb{R}^d\to\mathbb{R}$, see 
Refs.~\cite{Hansen-Goos2010edFMTlong,Wertheim_convex} for $k=2$ and $d=3$ and Appendix~\ref{sec:app_factorize}.
Here, we used the subspace determinant, $[\mathbf{n}_1,\dots,\mathbf{n}_k]\equiv \lvert\mathrm{det}(M)\rvert$~(see page 598 in \cite{Schneider_Weil}), where $M$ is the 
matrix $M$ whose elements are $M_{i,j}=\mathbf{n}_i\cdot\mathbf{e}^N_j$ for $1\leq i,j\leq k$ expressed in
some orthonormal basis $\mathbf{e}^N_j$ of the $k$-dimensional subspace spanned by the normal vectors.
With this definition, all $k$-body weight functions have the form
\begin{equation}
w^{[k]}\big(\mathcal B_1,\dots,\mathcal{B}_k;\mathbf{r})=
Q^{[k]}_d(\mathcal B_1,\dots,\mathcal{B}_k;\mathbf{r})
\prod_{i=1}^k w_{d-1}(\mathcal{B}_k,\mathbf{r}).
\label{eqn:w_contains_Q}
\end{equation}
Here, the functions $Q^{[k]}_d$ do not contain any distributions unlike $w^{[k]}$ and only depend on the local properties of the 
surfaces $\partial B_i$ at the intersection point $\mathbf{r}$. The $Q^{[k]}_d$ are summarized in Tbl.~\ref{tbl:kbody_w}.

\section{Fundamental mixed measure functional}\label{sec:ins}

Inserting the expression (\ref{eqn:normal_cone_w}) for $\chi[\mathrm{St}_n(\coords^n)]=\gamma(\mathrm{St}_n,\partial \mathrm{St}_n)$
into our approximation for the virial expansion of the excess free energy (\ref{eqn:Fexc_vir_app}) and recalling Eqn.~(\ref{eqn:c_n_value}) for $c_n$ 
, we 
find 
\begin{equation}
\mathcal{F}_\text{exc}=\int_{\mathbb{R}^d}\! \mathrm{d}\mathbf{r}\,\sum_{k=1}^d \Phi_{d}^{[k]},
\label{eqn:Fexc_FMMT}
\end{equation}
where 
\begin{align}
\Phi_{d}^{[k]}&\equiv n^{[k]}(\mathbf{r}) \sum_{n=k}^\infty \frac{k!}{n(n-1)} \binom{n}{k}
n_d(\mathbf{r})^{n-k} 
\nonumber\\&=
n^{[k]}(\mathbf{r}) \chi_k\big(n_d(\mathbf{r})\big)
\label{eqn:free_energy_density}
\end{align}
with
\[
\chi_k(\eta)=
\begin{cases}
\partial^k [(1-\eta)\log(1-\eta)+\eta]/\partial \eta^k & \text{for}\ \eta<1\ \text{and}\\
\infty & \text{for}\ \eta\geq 1.
\end{cases}
\]
Here, the weighted density $n_d$ is defined as usual, see Eqn.~(\ref{eqn:n_alpha_def}), while
the $k$-body weighted density is defined as
\[
n^{[k]}(\mathbf{r})\equiv \int_{\setofcoords^k}\! \mathrm{d}\coords^k\, w^{[k]}_0(\coords^k,\mathbf{r})\prod_{i=1}^k\rho(\coords_i).
\]
By integrating the $k$-body weight functions over all positions of the particles, we
obtain
mixed Minkowski volumes (or `fundamental mixed measures'), which are
generalizations of the fundamental measures to multiple bodies~\cite{Minkowski_Volumen}.
Therefore, the DFT with the functional (\ref{eqn:Fexc_FMMT}) could be called
`fundamental mixed measure theory' (FMMT). A FMMT functional containing only one-body and two-body weighted densities
has already been derived from the lowest order virial order and applied to spherocylinders~\cite{Wittmann_MM_smectic_letter}.

Note that the free energy is infinite if $n_d(\mathbf{r})> 1$ for any $\mathbf{r}$. This is 
a physical divergence, as $n_d(\mathbf{r})> 1$ implies that the point $\mathbf{r}$ lies inside more than one particle on average;
therefore, some particles must overlap at $\mathbf{r}$ and the free energy should indeed be infinite.

We will now compare the FMMT functional to the exact result in the zero dimensional limit.

\subsection{Comparison to the zero dimensional limit}\label{sec:zeroD}

Consider a mono-disperse system of hard particles in a quasi-zero-dimensional system, that is 
in a cavity that is so small that only one particle fits into the cavity. Alternatively,
in a multi-component system an artificial external potential can be considered,
which allows only particles of a single species to be inserted.
Denote the accessible domain of the particle by $\setofcoords$, the set of coordinates $\coords$ (positions and orientations) 
that the particle can have without extruding from the cavity. The set of accessible coordinates $\setofcoords$ is not necessarily 
connected as the cavity can have any shape.

The usual approach to obtain the free energy of this system starts with the grand canonical partition sum $\Xi$ that can be calculated exactly
for this quasi-zero-dimensional system,
$
\Xi=1+z \lvert \setofcoords\rvert
$, where $z=\exp(\beta \mu)/\mathcal{V}$ is the fugacity and $\lvert \setofcoords\rvert$ is the accessible hyper-volume.  
From this, the excess free energy can be calculated~\cite{Rosenfeld1996oldFMT},
\begin{equation}
\beta\mathcal{F}_\text{exc}=(1-x)\log(1-x) + x. \label{eqn:Fexc_0D_exact}
\end{equation}
where $x=\langle N\rangle$ is the average number of particles in this system,
which is less than one.

A second approach to obtain the excess free energy would be to perform a virial expansion for this system.
For all configurations that contribute to a Mayer diagram, that is, for all configurations $\coords^n\in\setofcoords^n$,
each particle always overlaps with all other particles by construction.
Therefore, each of the Mayer diagrams can be evaluated easily in this quasi-zero-dimensional system, as the Mayer bonds are always
$-1$ and the density profile $\rho(\coords)=\rho\equiv \langle N\rangle/\lvert \setofcoords\rvert$
is constant for all $\coords\in \setofcoords$ (and $\rho(\coords)=0$ if $\coords\notin\setofcoords$).
The resulting value of a diagram $g$ with $n$ nodes and $\lvert P(g)\rvert$ bonds is
\begin{equation}
g=\frac{1}{\lvert\text{Aut}(g)\rvert} \int_{\setofcoords^n}\!\mathrm{d}\coords^{n}\, (-1)^{\lvert P(g)\rvert} \rho^{n} = \frac{(-1)^{\lvert
P(g)\rvert}}{\lvert\text{Aut}(g)\rvert}x^{n}. \label{eqn:vir_0D_1}
\end{equation}
Therefore, the virial result for the excess free energy~(\ref{eqn:Fexc_vir_exp}) in this case is 
\begin{equation}
\beta F_\text{exc}=
-\sum_{n=2}^\infty \Bigg[ \sum_{g \in \mathcal{M}[n]} \frac{(-1)^{\lvert P(g)\rvert}}{\lvert\text{Aut}(g)\rvert} \Bigg] x^n. \label{eqn:vir_0D_2}
\end{equation}
As before, we use the combinatorial result~\cite{Ree_virial} $c_n\equiv-\sum_{g \in \mathcal{M}[n]} (-1)^{\lvert P(g)\rvert}/\lvert\text{Aut}(g)\rvert=1/[n(n-1)]$ 
to obtain
\begin{equation}
\beta F_\text{exc}=\sum_{n=2}^\infty \frac{x^n}{n(n-1)}=(1-x)\log(1-x) + x,\label{eqn:vir_0D_3}
\end{equation}
which, of course, is equal to the exact excess free energy (\ref{eqn:Fexc_0D_exact}), as obtained from the partition sum.

A functional can be obtained using this system~\cite{Rosenfeld1996oldFMT} by taking the zero-dimensional limit, $\setofcoords \to \lbrace \coords^{0\text{D}}_i\rbrace$,
where $\coords^{0\text{D}}_i$ for $1\leq i\leq M$ are the only accessible (discrete) states in the resulting `zero-dimensional' cavity.
In this limit, the density profile is simply a sum over delta functions and the functional can be constructed by demanding
that the excess free energy from the functional goes to the exact free energy for this system~(\ref{eqn:Fexc_0D_exact}) in the zero dimensional
limit. For details on the calculation, see the original
works for spheres~\cite{Rosenfeld1996oldFMT,Tarazona1997cavities} and the extension to anisometric particles~\cite{MM_Goetzke_db_FMT_app_B}.
In two and three dimensions, the excess free energy from the functional could not be reduced to the exact expression (\ref{eqn:Fexc_0D_exact}) for
cavities for which $\bigcap_{i=1}^M\mathcal{B}( \coords_i^\text{0D})$ was empty. These cavities were subsequently ignored.
The virial expansion route to the excess free energy in Eqns.~(\ref{eqn:vir_0D_1})--(\ref{eqn:vir_0D_3})
shows that (i) the resulting functional is equivalent to the one 
obtained in this work by performing a virial expansion and ignoring all $n$-particle configurations $\coords^n$ for which $\bigcap_{i=1}^M\mathcal{B}(
\coords_i^\text{0D})=\emptyset$ and (ii) the nontrivial combinatorial result $c_n=1/[n(n-1)]$ can actually be obtained by considering the quasi-zero-dimensional
system as the excess free energy~(\ref{eqn:vir_0D_3}) has to be exact in that system. 

It should be noted that the final form for the functional proposed in 
Ref.~\cite{Tarazona1997cavities} for hard spheres differs from the functional obtained in Sec.~\ref{sec:ins}, as additional approximations 
were performed in Ref.~\cite{Tarazona1997cavities} to obtain an efficient expression and the functional was rescaled to obtain a more
accurate result for the homogeneous fluid. We outline these approximations in Sec.~\ref{sec:exp_and_scl}.

\subsection{Direct correlation function}\label{sec:dir_corr}

One way to test the accuracy of the approximated functional is to compare the second direct correlation function 
\begin{equation}
c(\coords_1,\coords_2) = - \beta \frac{\delta^2 F_\text{exc}}{\delta \rho(\coords_1)\delta \rho(\coords_2)}\label{eqn:c2_def}
\end{equation}
with simulation results and established theories, which we will do in the following. Since previous results for $c(\coords_1,\coords_2)$
are only available
for the homogeneous and isotropic bulk fluid (isotropic phase), we restrict ourselves to a constant
density profile. In principle, we could obtain $c(\coords_1,\coords_2)$ by inserting the constant density profile
into Eqn. (\ref{eqn:c2_def}) with the functional (\ref{eqn:Fexc_FMMT}) and performing the many-particle
integrals explicitly; however, we chose a simpler route using the kinematic formula from integral geometry (see below).

We start with
Eqn.~(\ref{eqn:Fexc_vir_app}) for $\mathcal{F}_\text{exc}$ in which the $\rho(\coords_i)$ occur in an explicitly symmetric fashion, which 
makes it easier to perform the functional derivatives. Inserting Eqn.~(\ref{eqn:Fexc_vir_app}) in Eqn.~(\ref{eqn:c2_def}) and using Eqn.~(\ref{eqn:c_n_value}) for $c_n$, we find that 
\begin{multline}
c(\coords_1,\coords_2) = \\ -\sum_{n=2}^\infty \int_{\setofcoords^{n-2}}\!\mathrm{d}\coords^{n}_3\,
\chi\big[\textstyle\bigcap_{i=1}^n
\mathcal{B}(\coords_i) \big]\, \prod_{i=3}^n
\rho(\coords_i), \label{eqn:c_of_r1}
\end{multline}
where $\int_{\mathbb{V}^{n-k}}\mathrm{d}\coords_{k+1}^n = \int_{\mathbb{V}}\mathrm{d}\coords_{k+1} \cdots \int_{\mathbb{V}}\mathrm{d}\coords_n$ and the
prefactor $c_n=1/(n(n-1))$ is cancelled by the factors $n(n-1)$ that appear in the second functional derivative of 
$\prod_{i=1}^n
\rho(\coords_i)$.
From this expression, it can be seen that $c(\coords_1,\coords_2)$ only depends on the properties
of $\mathcal{B}_{1\cap 2} \equiv
\mathcal{B}(\coords_1)\cap\mathcal{B}(\coords_2)$ and not on the properties of the separate particles.
As a result, we can consider $\textstyle\bigcap_{i=1}^n
\mathcal{B}(\coords_i)$ as the intersection of a fixed particle $\mathcal{B}_0 \equiv \mathcal{B}_{1\cap 2}$ 
and $n'$ moving
particles $\mathcal{B}_i=\mathcal{B}(\coords_{i+2})$ for $1\leq i\leq n'$, where $n'=n-2$. 

To calculate $c(\coords_1,\coords_2)$ for the isotropic and homogeneous fluid,
we will use
the iterated kinematic integral formula \cite[Theorem 5.1.5]{Schneider_Weil} 
from integral geometry, 
which leads to Isihara's formula for the second virial coefficient~\cite{Isihara_B2}
when restricted to the Euler characteristic and two convex particles in three dimensions.
The iterated kinematic formula
is also valid for quite general~\cite{Schneider_int_geo_tools} classes of \emph{non}-convex bodies $\mathcal{B}_i$ and for other 
intrinsic volumes than the Euler characteristic; however we only require the formula for the Euler characteristic, which reads
\begin{multline}
\int_{ G_d^n} \mathrm{d}g^n \chi\big(\mathcal{B}_0 \cap g_1 \mathcal{B}_1 \cap \cdots \cap g_n \mathcal{B}_n)\\
= \sum_{\substack{i_0,\cdots,i_n=0 \\ i_0 + \cdots + i_n = n d}}^d C_{i_0,\cdots,i_n} v_{i_0}(\mathcal{B}_0) \cdots v_{i_n}(\mathcal{B}_n),
\label{eqn:iter_kin_form}
\end{multline}
where $\int_{G_d^n} \mathrm{d} g^n\equiv \int_{G_d} \mathrm{d}g_1 \cdots \int_{G_d} \mathrm{d}g_n$ denotes the $n$-fold integral over $G_d$,
the group of rigid body motions (translations and rotations) isomorphic to $\mathbb{R}^d \times \mathbb{SO}(d)$,
normalized such that $\int_{G_d} \mathrm{d} g\, w_d(g \mathcal{B},\mathbf{r})=v_d(\mathcal{B})$; also,
 $v_i(\mathcal{B}_j)$ is the $i$th intrinsic volume of body $\mathcal{B}_j$ (for $d=3$, $v_0$, $\pi v_1$, $2 v_2$ and $v_3$ are the Euler characteristic, the
integrated mean curvature,
the surface area and the volume respectively) and, finally, $C_{i_0,\cdots,i_n}$ is a prefactor,
\[
C_{i_0,\cdots,i_n}\equiv i_0! \kappa_{i_0} \prod_{j=1}^n \frac{i_j! \kappa_{i_j}}{d!\kappa_d}
\]
with $\kappa_i$ the volume of the $i$-dimensional unit ball $B_i$ (\emph{i.e.} the solid sphere with unit radius).
For bodies with a smooth boundary, the intrinsic volumes $v_i$ for $1\leq i\leq d-1$ can be calculated using~\cite[Page 607]{Schneider_Weil}: 
\begin{equation}
v_i(\mathcal{B})
\equiv  \frac{\binom{d}{i}}{d \kappa_{d-i}}\int_{\partial \mathcal{B}} \mathrm{d}^{d-1}\mathbf{r}\, H_{d-i-1}(\mathcal{B},\mathbf{r}),\label{eqn:intr_vol_calc}
\end{equation}
where $H_0=1$ and $H_{j}$ is the product of $j$ principal curvatures averaged over all combinations of $j$ principal directions:
\[
H_{j}(\mathcal{B},\mathbf{r}) =\binom{d-1}{j}^{-1} \sum_{i_1,\cdots,i_j=0}^{d-1}\prod_{k=1}^{j} \kappa_{i_j}(\mathcal{B},\mathbf{r}),
\]
Also, we denote the volume of a body $\mathcal{B}$ by
 $v_d(\mathcal{B})$.
In principle, it should be possible to prove the kinematic formula, which is outside
of the scope of this work, by inserting $\chi(\mathrm{St}_n)=\gamma(\mathrm{St}_n,\mathbb{R}^n)$ from 
Eqn.~(\ref{eqn:normal_cone_w}) and performing the integrals over $\coords_j$ for $1\leq j\leq n$ and $\mathbf{r}$.

While the iterated kinematic formula looks complicated for large $n$, it should be realized that due to the condition $i_0 + \cdots + i_n = n d$ $\Leftrightarrow$ 
$\sum_{j=1}^n (d-i_j)=i_0$ only at most $i_0$ of the $i_j$ for $1\leq j\leq n$ are unequal to $d$, such that the factors in $C_{i_0,\cdots,i_n}$ corresponding
to the remaining $i_j$ are unity. 

In order to write the direct correlation function~(\ref{eqn:c_of_r1}) for the bulk fluid
in the form of the iterated kinematic formula, we rewrite the combined integral and sum over the
(generalized) coordinate $\coords_j$ in Eqn.~(\ref{eqn:c_of_r1})
as
\begin{multline*}
\int_\mathbb{V} \mathrm{d}\coords_j\, \rho(\coords_j) \chi( \mathcal{B}(\coords_j)\cap\cdots) =\\
 \sum_{s_j=1}^M \int_{G_d} \mathrm{d}g_j\, \bar\rho_{s_j}\,
\chi(g_j \mathcal{B}^{(0)}_{s_j}\cap\cdots),
\end{multline*}
for a constant density profile $\rho(\coords)=\bar\rho_s/\lvert \mathbb{SO}(d)\rvert$,
where $M$ is the number of species in the system,
 $\mathcal{B}^{(0)}_s$ is the set of points inside a particle of species $s$ centered at the origin with identity orientation, 
(see also Sec.~\ref{sec:notation}) and
$\lvert \mathbb{SO}(d)\rvert = \int_{\mathbb{SO}(d)} \mathrm{d}\mathcal{R}$ the
volume of the group of rotations in Eqn.~(\ref{eqn:c_of_r1}).
Now we can apply the iterated kinematic formula, and subsequently simplify the resulting expression by
denoting the number of $i_j$'s equal to $\alpha$ by $N_\alpha$ for $0\leq j\leq n$ and $k=\sum_{\alpha=0}^{d-1} N_\alpha$ is the number of $i_j$'s unequal
to $d$. 
Also we define the scalar (one-body) weighted  densities
\[
\tilde{n}_{i}= \frac{i! \kappa_{i}}{d!\kappa_d}\, \sum_{s=1}^M v_i(\mathcal{B}^{(0)}_s)\bar\rho_s
\] 
(note that $\tilde{n}_d=\eta$ is the packing fraction), which are normalized differently than the $n_A$ from FMT for $d=3$.
With these definitions,
the direct correlation function~(\ref{eqn:c_of_r1}) becomes
\begin{align}
c(\coords_1,\coords_2) &=-\sum_{n'=0}^\infty 
\sum_{i_0=0}^d i_0! \kappa_{i0} \, v_{i_0}(\mathcal{B}_{1\cap 2}) \nonumber\\
&\!\!\!\!\!\!\!\!\!\sum_{\substack{
	N_0,N_1,\ldots,N_{d-1} \geq 0\\
	d N_0 + (d-1)N_1 + \cdots + 1 N_{d-1}=i_0
}}\!\!\!
\tilde{n}_0^{N_0} 
\tilde{n}_1^{N_1} 
\cdots
\tilde{n}_{d-1}^{N_{d-1}} \nonumber\\
&\qquad\qquad\sum_{N_d\geq 0}
\delta_{N_0+\cdots+N_d,n'}
\frac{n'!}{N_0! N_1! \cdots N_d!}
\eta^{N_d}\hspace*{-1.5em} \nonumber\\
&\!\!\!\!\!\!=
-\sum_{i_0=0}^d v_{i_0}(\mathcal{B}_{1\cap 2}) 
\sum_{k=0}^{i_0}
c^{[k]}_{i_0}(\tilde{n}_0,\cdot\!\cdot\!\cdot,\tilde{n}_{d-1})\, \chi_{k+2}(\eta)\quad
\label{eqn:c_2_final}
\end{align}
where we used that the number of combinations of $i_j$ for $0\leq i\leq n'$  that lead to the same $N_\alpha$
factors $\tilde{n}_\alpha$ for $0\leq \alpha\leq d$ is given by $n'!/(N_0!\cdots N_d!)$ and we defined
\[
c^{[k]}_{i_0}
= i_0! \kappa_{i0} \!\!\!\!\!
\sum_{\substack{
	N_0,N_1,\ldots,N_{d-1} \geq 0\\
	N_0 + N_1 + \cdots + N_{d-1}=k\\
	d N_0 + (d-1)N_1 + \cdots + 1 N_{d-1}=i_0
}}\!\!\!\!\!
\frac{\tilde{n}_0^{N_0} }{N_0!}
\cdots
\frac{\tilde{n}_{d-1}^{N_{d-1}} }{N_{d-1}!}.
\]
Note that the latter sum contains only very few terms for low $d$ (at most one for $d=2,3$).

Now we will compare to available expressions from different theories and simulation results for the direct correlation function to assess
the accuracy of the functional. 
The virial series up to first order in density for general shapes and general $d$ reads
\[
c_\text{exact}(\coords_1,\coords_2)= f_M(\coords_1,\coords_2) + \raisebox{-0.5em}{\scalebox{1.5}{\textdiaWWB}}
+\cdots
\]
where $\textdiaWWB$ denotes
\[
f_M(\coords_1,\coords_2)
\int_\mathbb{V} \mathrm{d}\coords_3\, \rho(\coords_3)\,
f_M(\coords_2,\coords_3)
f_M(\coords_3,\coords_1).
\]
We see that the lowest order is satisfied by the FMMT approximation for $c(\coords_1,\coords_2)$ as
$f_M(\coords_1,\coords_2)=-\chi\big(\,\mathcal{B}(\coords_1)\cap\mathcal{B}(\coords_2)\,\big)$ and $c^{[0]}_0=1$.
In order to consider the first order in density, we have to connect the integrals in \textdiaWWB\ to geometry.
It is relatively easy to see~\cite{Mulder2005spherozonotopes}\cite[Eqn. (46)]{Schneider_int_geo_tools} that 
$(\mathcal{B}(\coords_i) - \mathcal{R}\mathcal{B}^{(0)}_s)$ is the region excluded for the center of $\mathcal{R}\mathcal{B}^{(0)}_s$ by $\mathcal{B}(\coords_i)$,
where
$\mathcal{A}-\mathcal{B}=\lbrace\, a-b \,\vert\, a\in \mathcal{A}\text{ and }b\in \mathcal{B}\,\rbrace$ for two bodies $\mathcal{A}$ and $\mathcal{B}$.
Therefore, the positional integral in $-\textdiaWWB$, whose integrand is nonzero if $\mathcal{R}_3\mathcal{B}^{(0)}_s+\mathbf{r}_3$
 overlaps with both $\mathcal{B}(\coords_1)$ and
$\mathcal{B}(\coords_2)$,
 can be written as the volume of $(\mathcal{B}(\coords_1) - \mathcal{R}_3\mathcal{B}^{(0)}_s) \cap
(\mathcal{B}(\coords_2) - \mathcal{R}_3\mathcal{B}^{(0)}_s)$~\cite{Rosenfeld1988SFP}.
The geometry of this region is different from the excluded region of $\mathcal{R}_3\mathcal{B}^{(0)}_s$ and $\mathcal{B}_{1\cap 2}$, which is
the corresponding result from FMMT, due to lost cases.
As a result, no amount of rescaling can fix this difference once and for all for general (mixtures of) shapes. 
Nevertheless, for particular shapes (see Sec.~\ref{sec:exp_and_scl}), the difference may be small or indeed zero.

\begin{figure}[b]
\vspace*{-0.5em}
\includegraphics[width=0.99\columnwidth]{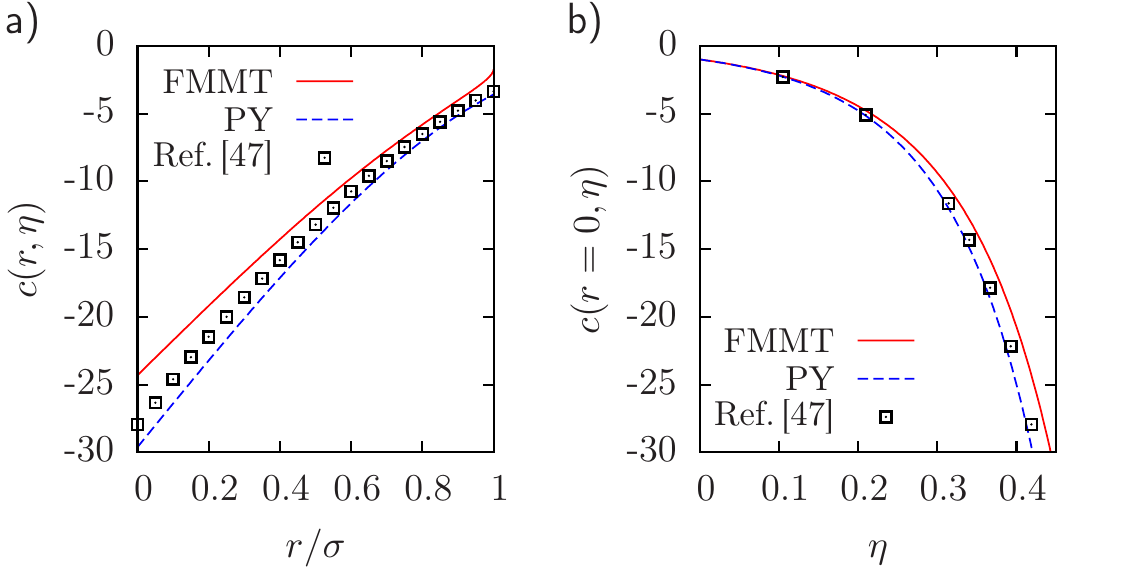}

\caption{
(a) The direct correlation function from FMMT, see Eqn.~(\ref{eqn:c_2_final}), at $6\eta/\pi=0.8$ compared to the Percus-Yevick $c(r,\eta)$ (PY) and simulation results by
Groot, Eerden and Faber~\cite{Groot_dir_corr_HS_sims}. (b) The behavior of the FMMT direct correlation function with varying $\eta$ at $r=0$, where lost cases do
not contribute to first order in $\eta$,
is compared with the Percus-Yevick result. As the lost cases do not contribute here, the first order in $\eta$ of the FMMT $c(r,\eta)$ is exact
(as is the first order PY result).
\label{fig:dir_corr}}
\end{figure}

In order to make this difference in geometries more explicit and examine effects of higher densities,
we turn to the bulk hard sphere fluid and compare to the Percus-Yevick (PY) direct correlation function~\cite{Wertheim_PY_HS}.
The FMMT $c(\coords_1,\coords_2)$ was calculated using Eqn.~(\ref{eqn:c_2_final}) and the intrinsic volumes $v_i(\mathcal{B}_{1\cap 2})$ of
the intersection of two spheres using Eqn.~(\ref{eqn:intr_vol_calc}) and $v_i(\mathcal{B}_{1\cap 2})= \lim_{\epsilon\to 0} v_i\big((\mathcal{B}_{1\cap 2})_\epsilon\big)$, see
Eqn.~(\ref{eqn:parallel_body}) for the definition of the parallel body $(\cdot)_\epsilon$ and Refs.~\cite{MM_Goetzke_db_FMT,auto_MM_polyhedra} 
for a similar procedure.
For hard spheres, all $v_i(\mathcal{B}_{1\cap 2})$ are polynomials in the distance $r$ between the centers
of $\mathcal{B}(\coords_1)$ and $\mathcal{B}(\coords_2)$ except for $v_1(\mathcal{B}_{1\cap 2})$, which is proportional to the 
mean half width or the integrated mean curvature and contains a term proportional to $\arcsin(r/\sigma)\sqrt{\sigma-r^2}$, where $\sigma$ 
is the hard sphere diameter. However, the first order exact contribution $-\textdiaWWB$ is the volume of the intersection
between two spheres with diameter $2\sigma$ at a center-to-center distance $r$, which is a polynomial in $r$ only.
So we again see that rescaling will not make the FMMT and exact first order contributions in $c(r,\eta)$ agree for all $r$ and $\eta$.
Nevertheless, the difference between the exact and the FMMT approach might still be numerically small in practice. To access
this difference, we compare the direct correlation function from FMMT to the ones from PY and simulations in Fig.~\ref{fig:dir_corr}. We see that
the deviation of FMMT from the simulation results~\cite{Groot_dir_corr_HS_sims} is larger than for PY; however, the FMMT
direct correlation function never performs more than an order of magnitude worse than the PY $c(\coords_1,\coords_2)$. 
We also show the three direct correlation
functions as a function of $\eta$ at $r=0$ (where lost cases do not contribute to the triangle diagram) to show that the lost cases in \textdiaWWB\
are not the only cause for the difference between the PY and FMMT predictions for $c(r,\eta)$.

We will now review the methods that have been used in the literature to overcome these difficulties for hard spheres.

\subsection{Expansion and rescaling}
\label{sec:exp_and_scl}

First, the equation
of state for the homogeneous fluid from the functional from Sec.~\ref{sec:ins} is not very accurate. Because of the neglected lost cases,
the density expansion of the FMT free energy is already inexact at the third virial order. 
This is especially pronounced for thin rods, where a significant contribution of
the exact triangle diagram is due to lost cases where the three particles are nearly coplanar and the three regions of pairwise intersection are
well-separated.
This effect of lost cases becomes especially problematic in two dimensions, where any FMT-like functional would incorrectly predict a vanishing 
third virial coefficient for infinitely thin needles. This same problem would also occur in a highly ordered uni-axial 
nematic phase of biaxial platelets (when the particle is very thin along one of its axes)~\footnote{
The same prediction results from the Zwanzig model in FMT, where each of the particle axes lies along one of the
Cartesian axes; however, the \emph{exact} third order virial coefficient also vanishes in that case.
}. 
On the other hand, the FMMT functional for hard parallel cubes, which is equal to Cuesta's functional~\cite{Cuesta_paraPRL2},
is exact at the third virial order because there are no lost cases for hard cubes. In fact, this also holds independently of the edge-lengths
for (mixtures of) other single-orientation parallelepipeds and their $d$-dimensional generalizations with $2d$ facets,
provided that each facet of each species is perpendicular to
one of $d$ linearly independent directions
$\mathbf{u}_i$~\footnote{Define a non-orthogonal basis $\mathbf{e}_i$ of $\mathbb{R}^d$ by $\mathbf{e}_i\cdot\mathbf{n}_j=0$ 
for all $i\neq j$.
All positions can be uniquely decomposed in components $r_j$:
$\mathbf{r}=\sum_{i=1}^d r_j \mathbf{e}_i$
. For each particle $i$ in a configuration, $d$ intervals $[a^{(i)}_j,b^{(i)}_j]$ can be defined such that a point $\mathbf{r}$ 
lies in the interior of particle $i$ if and only if $a^{(i)}_j\leq r_j\leq b^{(i)}_j$ for all $1\leq j\leq d$. 
By the argument in footnote~\cite{Note6}, there are no lost cases.
}.

In Ref.~\cite{Tarazona1997cavities},
as well as in the original derivation of FMT by Rosenfeld~\cite{Rosenfeld1989FMTPRL},
the $\Phi_{3}^{[3]}$ term in the functional for hard spheres was rescaled
to obtain the exact third virial coefficient. For anisotropic particles, the prefactor of $\Phi_{3}^{[3]}$ is either kept equal to that
of spheres~\cite{Hansen-Goos2010edFMTlong} or it is modified~\cite{MM_Zimmermann_rcubes}
such that the correct third virial coefficient for anisotropic particles is obtained
(the third virial coefficient has to be calculated numerically in general).

Secondly, the generalized weighted densities $n^{[k]}_0$ contain $k$ integrals over the particles' coordinates,
such that calculating these directly is computationally involved for $k>1$ and $d>2$~%
\footnote{
Evaluating 
$n^{[k]}(\mathbf{r})$ once for every position for a density profile
where all of the $d(d-1)/2$ orientational and $d$ positional degrees
of freedom are discretized on a grid of size $G$ for all $M$ species requires $ G^d\, \big(M\,G^{d-1+d(d-1)/2}\big)^k$
evaluations of $Q^{[k]}_d$;
for $d=k=3$, $M=1$ and a grid size of $G=10^n$ this corresponds to $10^{13 n}$ evaluations.
}%
. An efficient functional can be obtained if
the generalized weighted densities for $k\geq 2$ are expanded
in products of $k$ single-particle weight functions. 
Using a similar calculation as 
Wertheim's deconvolution of the Mayer bond~\cite{Wertheim_convex},
the $n^{[2]}$ weighted density in three dimensions can be expanded in either
tensor weighted densities (that is weighted densities that transform as tensors under a rotation of the basis vectors)
or weighted densities that transform as the spherical harmonics under a rotation of the basis vectors.

\begin{figure}
\includegraphics[width=0.6\columnwidth]{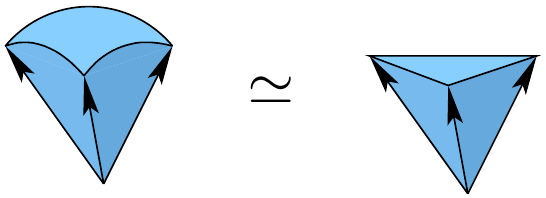}

\caption{ The geometrical consideration behind the approximation (\ref{eqn:third_term_app}) 
for the third term in the three dimensional FMT functional.
\label{fig:third_term_app}}
\end{figure}

For the kernel of the third term, $Q^{[3]}_3$, Tarazona and Rosenfeld's approximation~\cite{Tarazona1997cavities} for hard spheres can be reinterpreted
as the geometrical approximation shown in Fig.~\ref{fig:third_term_app}. In this approximation, we first write the
area of the spherical triangle spanned by three normal vectors $\mathbf{n}_i$ at a point $\mathbf{p}$ on a triple surface intersection
as three times the 
volume of the corresponding section of a unit ball. Subsequently, we replace the volume of this section of the unit ball by 
the volume of a tetrahedron that has the normal vectors as three of its edges, $\frac{1}{6}\lvert \mathbf{n}_1\cdot (\mathbf{n}_2\times\mathbf{n}_3)\rvert$.
The final form for the approximation then reads $\sigma_{2}\big(N^+(\mathrm{St}_n,\mathbf{p})\big)\simeq \frac{1}{2}\lvert \mathbf{n}_1 \cdot
(\mathbf{n}_2\times\mathbf{n}_3)\rvert$, such that the kernel becomes
\begin{equation}
Q^{[3]}_3 \simeq \frac{1}{48\pi} [\mathbf{n}_1 \cdot (\mathbf{n}_2\times\mathbf{n}_3)]^2 \label{eqn:third_term_app}
\end{equation}
for general shapes. This approximation is exact in the limit that the density profile goes to a single infinitely sharp peak
centered around some position and orientation.

Approximations for $n^{[k]}$ have to be formulated with some care, because otherwise 
some of the important properties of the generalized weight function might be lost.
For instance, the exact $Q^{[k]}_d$ for $k\geq 2$ vanishes 
if one of the particles $i$ is moved on top of another particle $j$,
$\lim_{\coords_i\to\coords_j}
Q^{[k]}_d(\coords^k,\mathbf{r})=0$, due to the prefactor
$[\mathbf{n}_1,\cdots,\mathbf{n}_k]$
 and, at least for $k=d=3$,
it is important~\cite{Tarazona1997cavities,Tarazona2000FMT}
that the approximation for $Q^{[3]}_3$ also vanishes in this limit.
Failing to take this condition into account causes a negative divergence in the zero dimensional limit for hard spheres~\cite{Rosenfeld1996oldFMT}
and causes the crystal to be unstable with respect to the fluid for the whole density range.

After the expansion was performed for hard spheres, the third term in the functional for $d=3$ was rescaled to obtain the correct third virial
coefficient for hard spheres~\cite{Tarazona1997cavities}, which leads to the PY equation of state (via the compressibility route).
The kernel $Q_3^{[3]}$ has also been modified by Tarazona~\cite{Tarazona2000FMT} to obtain the exact triangle diagram in 
the direct correlation function for the homogeneous fluid of hard spheres.
The latter calculation results, without further modifications, in the Percus-Yevick direct correlation function for spheres~\cite{Wertheim_PY_HS}.

Finally, we have made approximations beyond just neglecting the lost cases: we ignored all Ree-Hoover diagrams 
other than $(\emptyset)_n$. These can be taken into account approximately for the homogeneous fluid by adapting the functional
to fit some equation of state (EOS), which is then an input for the theory rather than a result.
The modifications proposed by Roth \emph{et al}~\cite{Roth2002,Hansen-Goos2006WBII},
multiplying each term $\Phi_{d}^{[k]}$ 
by a certain function of $n_d(\mathbf{r})$ only, lead to the so-called White Bear II functional, which is still accurate for the crystal and
results by construction in the
highly accurate Carnahan Starling equation of state for the homogeneous fluid of spheres.
Note, that semi-empirical modifications that improve the EOS of the fluid for specific shapes
do not necessarily improve the results for strongly inhomogeneous and/or anisotropic density profiles
and that the White-Bear II functional is only accurate for moderately non-spherical shapes even in the homogeneous
case~\cite{Rosenfeld1988SFP,Wittmann_MM_smectic}.

\section{Summary and discussion}
\label{sec:summary}

We have derived a density functional from an approximated and resummed virial expansion
for hard particles with arbitrary convex shapes in $1\leq d\leq 3$ dimensions. 
While all Mayer diagrams were considered, we approximated each diagram by neglecting those configurations
for which the intersection between all particles was empty.
This is the only approximation in our derivation. All approximated $n$-particle diagrams 
become proportional to the same integral, while the sum of the prefactors could be obtained 
by comparing to Ree and  Hoover's resummation of the Mayer diagrams~\cite{Ree_virial}.
Using the 
geometry of the $n$-particle intersection, we wrote the approximated Mayer diagrams
in terms of generalized, $k$-body weight functions and resummed the series to a closed form
containing $d$ terms in $1\leq d\leq 3$ spatial dimensions.
The resulting functional equals the fundamenal measure functional (0D-FMT) that was earlier derived from consideration of 
an extremely confined geometry (the `zero-dimensional limit'), 
which tells us that the virial series is actually contained, in an approximate sense,
in the previously proposed FMT functionals.

The geometric formulation of the resulting functional has the advantage 
that results from integral geometry can be directly transfered to 0D-FMT,
which we used to calculate the direct correlation function for constant densities. 
We showed that the direct correlation function thus obtained has a different geometrical origin than
the exact virial expansion for general shapes already at the $\propto \rho$ term.
A similar fundamental difference in form is also found when comparing the 0D-FMT result
to the established Percus-Yevick (PY) result for hard spheres at finite density, although, in practice, 
the accuracy of 0D-FMT turns out to be mostly comparable to that of PY for the direct correlation function at high densities.

The generalized weight functions in 0D-FMT contain integrals over the coordinates 
of $1\leq k\leq d$ particles. To simplify the generalized
weight functions, they can be expanded into one-body weight functions. We briefly
reviewed the possibilities to perform this expansion as proposed in Refs.~\cite{Tarazona1997cavities,Tarazona2000FMT,Hansen-Goos2010edFMTlong,Wertheim_convex}
and reiterated the conditions for this expansion to yield a functional for a given $d$ that correctly reduces to the functional for $d'<d$ 
when applied to a system under strong confinement.
Correct behavior under such dimensional reduction has turned out to be important for the crystal of hard
spheres~\cite{Rosenfeld1996oldFMT,Tarazona2000FMT}.

The success of FMT for spheres is perhaps unexpected considering the severity of our approximation
and the ones that were made afterwards~\cite{Tarazona1997cavities,Tarazona2000FMT}.
The effect of our approximation becomes more clear if Ree and Hoover's resummation of the Mayer diagrams into a sum of other
types of diagrams~\cite{Ree_virial} is considered. 
We showed that neglecting all configurations with an empty $\mathrm{St}_n$ in the $n$-particle Mayer diagrams
is equivalent to the following two approximations on the Ree-Hoover diagrams:
First, we neglect all $n$-particle Ree-Hoover diagrams but
$(\emptyset)_n$, the
Ree-Hoover diagram where each particle is required to overlap with all other particles. 
Subsequently, the $(\emptyset)_n$ diagram is approximated by neglecting the
`lost cases'~\cite{Tarazona1997cavities}, which are configurations for which each particle overlaps with all other particles,
but there is no common region of overlap ($\mathrm{St}_n=\emptyset$).
The former approximation would lead to an overestimation of the free energy,
at least for the bulk fluid of hard spheres~\cite{Ree_virial,Ree_7vir}, as the
neglected $e_\text{RH}$-bonded Ree-Hoover diagrams sum up to a net negative contribution.
The latter approximation, 
neglecting the lost cases, leads to an underestimation of the $(\emptyset)_n$ diagram
(both the exact and approximated $(\emptyset)_n$ diagrams
are positive if the combinatorial prefactors are included) even after rescaling
to obtain the correct third order diagram; therefore, the approximation lowers
the free energy.
The effect of these two approximations cancels partially for hard spheres and probably also for
other shapes, which might explain
FMT's success. As mentioned in Sec.~\ref{sec:exp_and_scl}, particles like hard cubes 
have no lost cases, such that no partial cancellation occurs.
Nevertheless, 
the phase behavior of hard parallel cubes is described reasonably well by Cuesta and Mart\'inez-Rat\'on's FMT functional~\cite{Cuesta1997I,martinez-raton1999}.

We note that our functional for $d=1$ agrees with the exact functional for inhomogeneous hard-rod mixtures that was derived by Vanderlick, Davis and
Percus~\cite{Vanderlick_hard_rods}.  In $d=1$ dimensions, there are no lost cases for hard rods~\footnote{
Sort the $k$ line intervals, $[a_i,b_i]$, such that $a_1<a_2<\ldots<a_k$. 
If all of the intervals overlap pairwise, then $a_k < b_j$ for $j<k$.
Since also $a_k>a_j$ for $j<k$, this implies that all intervals contain $a_k$,
such that $\bigcap_{i=1}^k [a_i,b_i]\neq \emptyset$.
}, so the $(\emptyset_n)$ diagram is exactly contained in our functional.
Therefore, we have proven that only the $(\emptyset_n)$ Ree-Hoover
diagrams contribute for hard rods for $d=1$ for any density profile, which was shown
before for homogeneous systems and conjectured to be true for inhomogeneous density profiles~\cite{Ree_virial}. 
As mentioned before, the success of a functional for crystallization is highly dependent on the behavior under dimensional reduction.
As our functional and the other modern FMT functionals for $d>1$ by construction reduce to the $d=1$ case under confinement
in the narrowest possible (straight) channel,
this might also explain some of the successes of the functional.

Finally, one of the requirements that allowed Rosenfeld to derive FMT is that FMT obeys the scaled particle relation, \emph{i.e.}
the excess chemical potential for adding a large particle to the mixture becomes equal to 
work against pressure required to clear a region of the size of the particle in the limit when the particle becomes macroscopically large.
The version from the zero-dimensional limit and the other versions of FMT exhibit the correct scaled
particle limit, which probably also adds to the accuracy of the functional.

Now, we are able to discuss possible improvements beyond the current functional. The first and most obvious improvement would be to include
the `lost cases'. Wertheim has shown how \scalebox{0.6}{$\triadia$}, the lowest-order Mayer diagram that suffers from lost cases, 
can be written in terms of two-center weighted densities, that is, generalized weighted densities that depend on two positions.
Future improvements should probably include these two-center weighted densities, as the lost cases 
cannot be recovered for a general density profile if only one-center weighted densities are used.

\begin{figure}
\includegraphics[width=0.9\columnwidth]{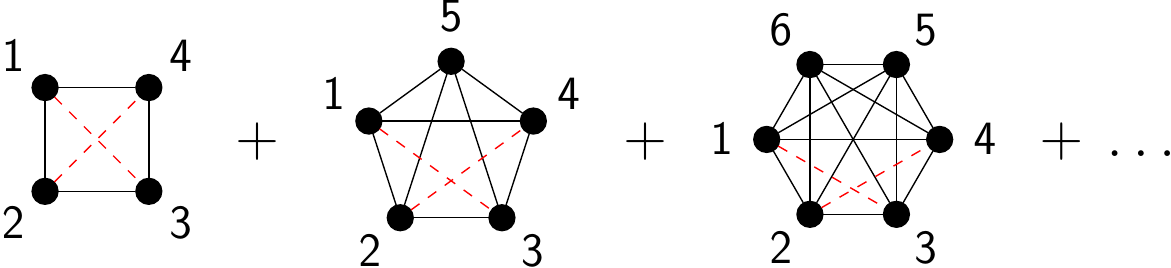}

\caption{
 The next diagrams in the Ree-Hoover resummation, if the diagrams are added in order of increasing
number of $e_\text{RH}$-bonds (dashed lines) or, equivalently, decreasing number of $f_M$ bonds (solid lines).
The sign and combinatorial prefactor~\cite{Ree_virial} are contained in the diagram.
\label{fig:next_RH_diagrams}}
\end{figure}

Secondly, other Ree-Hoover diagrams might be added to the functional, and (after suitable approximation)
be written in terms of two-center weighted densities. For hard spheres, for example, Ree and Hoover~\cite{Ree_7vir} noted 
that the $n$-particle diagrams in the virial expansion for the homogeneous fluid could be approximated reasonably well for $n\leq7$ by including
only the $(\emptyset)_n$ diagram and
the diagram with two $e_\text{RH}$ bonds, see Fig.~\ref{fig:next_RH_diagrams}.
After suitable labeling, the integrand in the latter diagram is nonzero if all pairs overlap except $(1,3)$ and $(2,4)$.
In Ref.~\cite{arXiv_Korden_DFT_N_dims}, several functionals containing multiple-center weighted densities are derived from the
Ree-Hoover diagrams.

Finally, we could consider adding the Mayer ring diagrams, in which
each particle $i$ is only connected by a Mayer bond to two other particles, for example 
\scalebox{0.6}{$\triadia$} and \scalebox{0.6}{$\fourdiaone$} are the ring diagrams with three and four particles, respectively.
Due to the loose connectivity, this diagram can be evaluated with relative
ease for the homogeneous fluid~\cite{Montroll_Mayer_chain_diagrams}
and can be expressed in terms of Wertheim's two-center weighted densities~\cite{Wertheim_convex}. It would be interesting to consider replacing 
our approximations for the ring diagrams (and similar loosely connected diagrams) by their exact values and see if this leads to an improvement
of the functional.

\begin{acknowledgments}
The idea for this paper originated during a visit of MM and SK to Martin Oettel in 
T\"ubingen. We would like to thank Martin Oettel, Roland Roth and others present
at the time for fruitful discussions. Financial support by the DFG
under grant Me1361/12 as part of the Research Unit 'Geometry and Physics of Spatial Random Systems'
is gratefully acknowledged by KM, while SK was supported by 
the Cluster of Excellence ``Tailor-Made Fuels from Biomass'', funded by
the Excellence Initiative of the German federal and state governments.
\end{acknowledgments}

\appendix

\section{Appendix: intersections of delta shells}
\label{sec:app_factorize}

In this section, we will derive Eqn.~(\ref{eqn:simpl_intersect}), which is a generalization
of the results for $k=2$~\cite{Hansen-Goos2010edFMTlong} and $k=3$~\cite{Korden_deriving_FMT} in $d=3$ dimensions.
Consider the parallel body of the intersection between $k$-surfaces $\partial \mathcal B_i$,
$(\inters_k)_\epsilon\equiv (\partial\mathcal B_1\cap\cdots\cap\partial\mathcal B_k)_\epsilon$ and
let $\mathcal{N}(\mathbf{p})$ be the $k$-dimensional subspace spanned
by the normal vectors of the intersecting surface at $\mathbf{p}$.
The integrand on the right hand side of Eqn.~(\ref{eqn:simpl_intersect}) is
zero if $\mathbf{r}\not\in(\inters_k)_\epsilon$ for some $\epsilon$, such that the right hand side of Eqn.~(\ref{eqn:simpl_intersect}) becomes
\[
\int_{(\inters_k)_\epsilon}\! \mathrm{d}\mathbf{r}\, h(\mathbf{r})\prod_{i=1}^k w_{d-1}(\mathcal B_i,\mathbf{r}),
\]
where $h(\mathbf{r})=f(\mathbf{r})[\mathbf{n}_1,\dots,\mathbf{n}_k]$.
Also,
if $\epsilon$ is small enough, we can locally approximate $(\inters_k)_\epsilon$ as
the parallel body of a flat $d-k$ dimensional plane, which allows us to 
approximate $(\inters_k)_\epsilon$ as
$\lbrace \mathbf{p}+\mathbf{x} | \mathbf{p}\in \inters_k,\; \mathbf{x}\in \mathcal{E}_\epsilon(\mathbf{p}) \rbrace$,
where $\mathcal{E}_\epsilon(\mathbf{p})$
consists of those elements $\mathbf{r}$ of $\mathcal{N}(\mathbf{p})$ such that $\lvert \mathbf{r}\rvert <\epsilon$. 
With this approximation (which is exact in the limit $\epsilon \to 0$), we can write the
right hand side of Eqn.~(\ref{eqn:simpl_intersect}) as 
\[
\lim_{\epsilon\to0}\int_{\inters_k}\! \mathrm{d}^{d-k}\mathbf{p}\, \int_{\mathcal{E}_\epsilon(\mathbf{p})}\! \mathrm{d}\mathbf{x}\,
h(\mathbf{p}+\mathbf{x})\prod_{i=1}^k \delta\big(\mathbf{x}\cdot \mathbf{n}_i(\mathbf{p})\big),
\]
where we locally approximated $\inters_k$ as
the intersection of $k$ flat surfaces with normal vectors $\mathbf{n}_i(\mathbf{p})$. Now 
we 
parametrize $\mathbf{x}=\sum_{i=1}^{k} x_i\mathbf{e}^N_i$, where the $\mathbf{e}^N_i$ form an 
orthonormal basis of $\mathcal{N}(\mathbf{p})$ and perform the variable transformation $(x_1,\dots,x_{k})\to \mathbf{t}$, where $t_i=\mathbf{x}\cdot
\mathbf{n}_i(\mathbf{p})$ is the argument of the $i$th delta function in the expression above for $1\leq i\leq k$.
The Jacobian matrix of the \emph{inverse} transformation $\mathbf{t} \to (x_1,\dots,x_{k})$ has components 
$M_{ij}=\partial t_i/\partial x_j=\mathbf{n}_i(\mathbf{p})\cdot\mathbf{e}^N_j$, such that the Jacobian determinant
of $(x_1,\dots,x_{k})\to \mathbf{t}$ equals $1/\lvert\mathrm{det}(M)\lvert=1/[\mathbf{n}_1,\dots,\mathbf{n}_k]$.
After performing
the variable substitution $(x_1,\dots,x_{k})\to \mathbf{t}$
and the integrals over the $t_i$, we obtain the left hand side
of Eqn.~(\ref{eqn:simpl_intersect}), which completes the proof.

\end{document}